\documentclass[aps,pra,reprint,groupedaddress,showpacs]{revtex4-1}
\usepackage{graphicx}
\usepackage{latexsym,amssymb,amsmath}
\usepackage{bm}
\usepackage[caption=false]{subfig}
\usepackage{xcolor}
\usepackage{mathrsfs}
\usepackage[colorlinks=true,
urlcolor=blue,
linkcolor=blue,
citecolor=blue
]{hyperref}

\begin{document}
\title{Terahertz topological plasmon polaritons for robust temperature sensing }

\author{B. X. Wang}
\author{C. Y. Zhao}
\email{changying.zhao@sjtu.edu.cn}

\affiliation{Institute of Engineering Thermophysics, School of Mechanical Engineering, Shanghai Jiao Tong University, Shanghai, 200240, China}
\date{\today}T

\begin{abstract}
We theoretically investigate the application of topological plasmon polaritons (TPPs) to temperature sensing for the first time. Based on an analogy of the topological edge states in the Su-Schrieffer-Heeger model, TPPs are realized in a one-dimensional intrinsic indium antimonide (InSb) microsphere chain. The existence of TPPs is demonstrated by analyzing the topology of the photonic band structures and the eigenmode distribution. By exploiting the temperature dependence of the permittivity of InSb in the terahertz range, the resonance frequency of the TPPs can be largely tuned by the temperature. Moreover, it is shown that the temperature sensitivity of the TPP resonance frequency can be as high as $0.0264~\mathrm{THz/K}$ at room temperature, leading to a figure of merit over 150. By calculating the LDOS near the chain, we further demonstrate that the temperature sensitivity of TPPs is experimentally detectable via near-field probing techniques. This sensitivity is robust since TPPs are highly protected modes immune to disorder and can achieve a strong confinement of radiation. We envisage these TPPs can be utilized as promising candidates for robust and enhanced temperature sensing.
\end{abstract}

\maketitle
Topological phases of matter can support robust edge states immune against scattering from disorder and imperfections, which have received a great deal of attention in recent years and been demonstrated for electronic \cite{hasanRMP2010}, electromagnetic \cite{ozawa2018topological}, acoustic \cite{heNaturephys2016}, cold atomic \cite{atalaNaturephys2013} and mechanical \cite{susstrunkScience2015} systems. In particular, since topological photonic systems can hold topologically protected optical modes \cite{luNPhoton2014,khanikaevNPhoton2017,riderJAP2019,xieOE2018}, they provide great opportunities for achieving precise, robust and local control of light, which facilitate high-efficiency photonic devices such as unidirectional waveguides \cite{poliNComms2015}, optical isolators \cite{el-GanainyOL2015} and topological lasers \cite{stjeanNaturephoton2017,partoPRL2018,zhaoNaturecomms2018}. Notably, as a unique combination of topological protection and strong light confinement due to plasmonic excitations, topological plasmon polaritons (TPPs) are arguably among the most promising approaches to robust and deep-subwavelength scale light-matter interactions and have therefore attracted growing attention in the last a few years \cite{lingOE2015,downingPRB2017,pocockArxiv2017,downing2018topological,pocockNanophoton2019}. For instance, the modal wavelength of topologically bounded plasmonic modes in multilayered graphene systems can be squeezed as small as 1/70 of the incident wavelength \cite{xuAppsci2019}. Low-power-consumption and highly-integrated four-wave mixing processes can also be engineered through the TPP modes in graphene metasurfaces \cite{you2019four-wave}. As a result, the topological protection of the spectral and spatial position of TPPs with enhanced light-matter interactions indeed provides a novel route to robust sensing. However, to the best of our knowledge, there have been very few works concerning topological optical states for sensing applications so far.

\begin{figure}[htbp]
	\centering
	\subfloat{
		\includegraphics[width=0.8\linewidth]{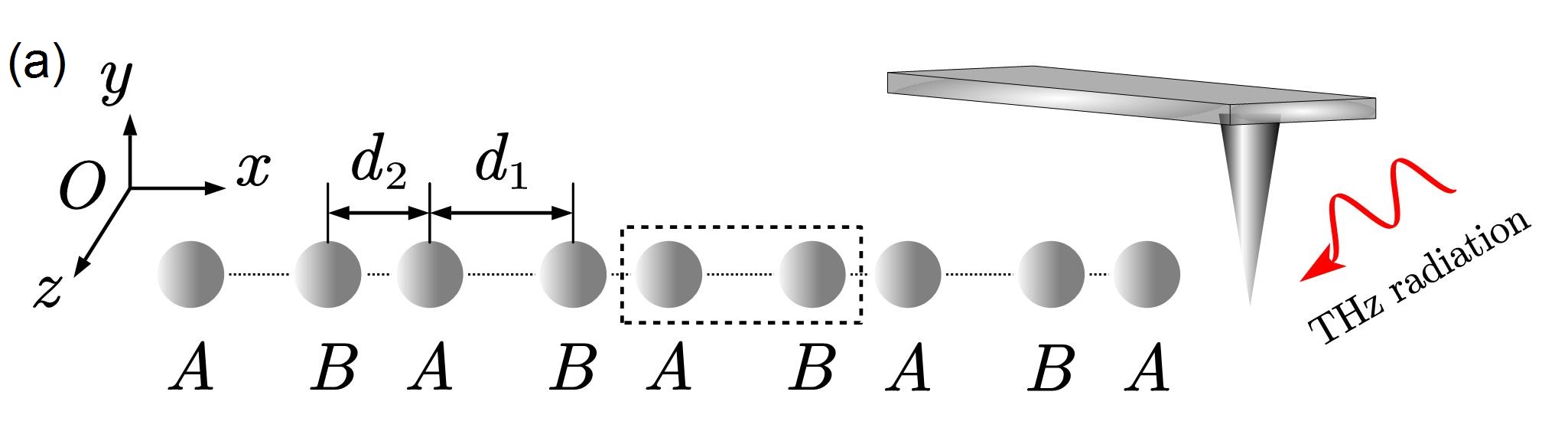}\label{schematic_chain}
	}
	\hspace{0.01in}
	\subfloat{
		\includegraphics[width=0.46\linewidth]{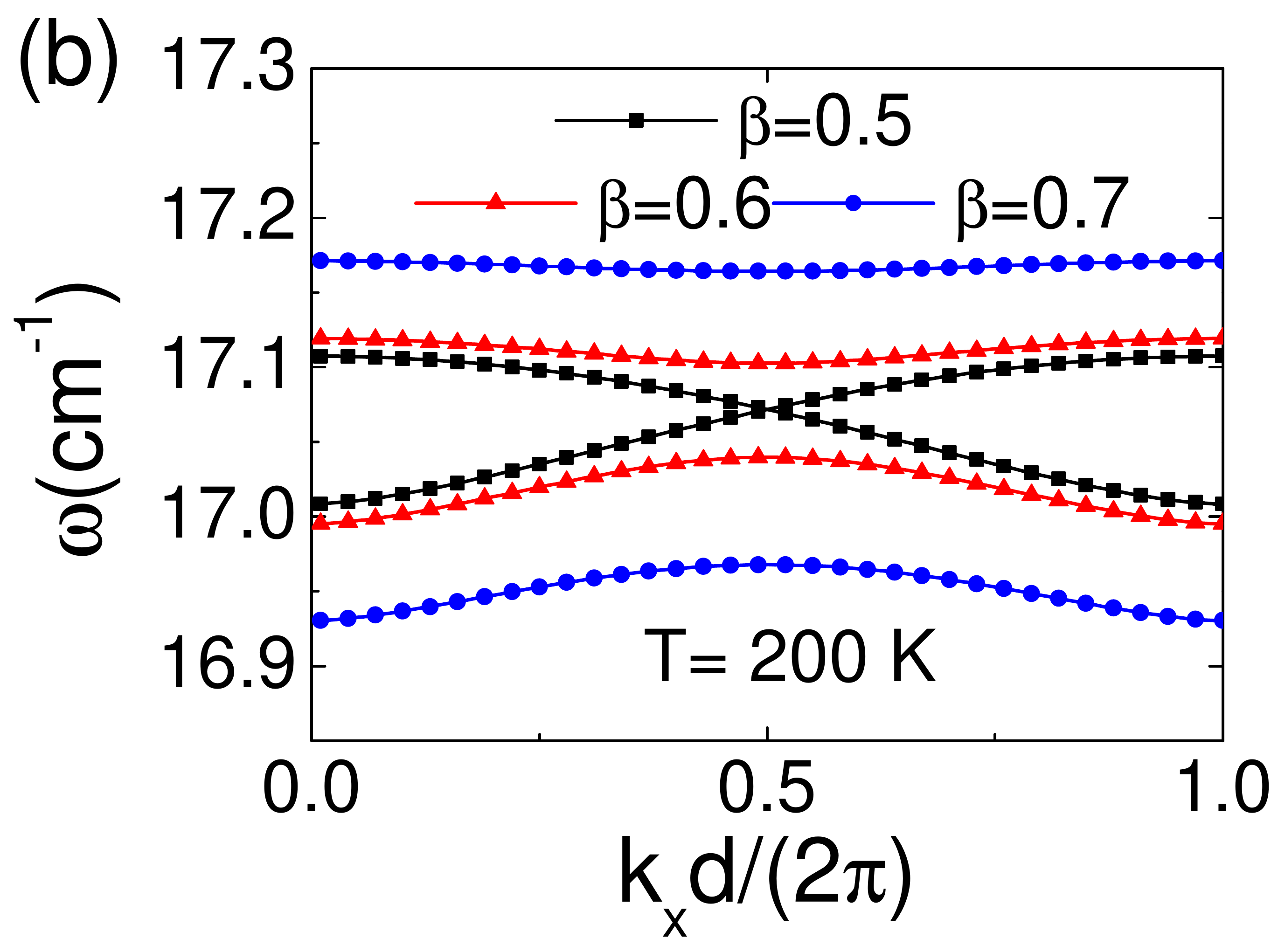}\label{bandstructurelongd10}
	}
	\hspace{0.01in}
	\subfloat{
		\includegraphics[width=0.46\linewidth]{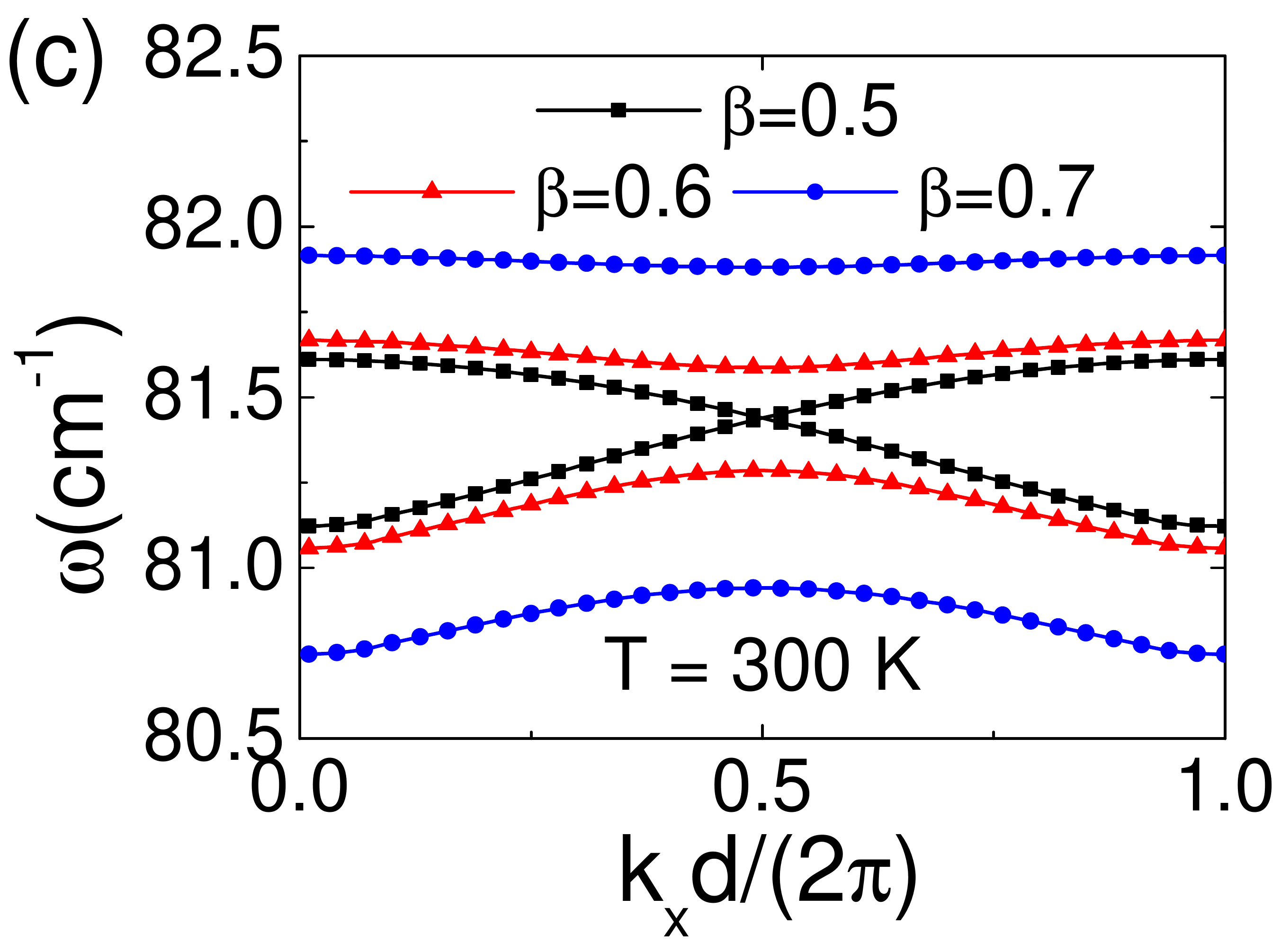}\label{bandstructurelongd10T300}
	}
	\caption{(a) Schematic of the dimerized InSb microsphere chain with a SNOM tip nearby and THz incident radiation. (b-c) Real parts of the longitudinal band structures of a dimerized InSb microsphere chain under a temperature of (b) 200 K and (c) 300 K with different $\beta$.}\label{figcda}
	
\end{figure}

In this work, we theoretically explore the possibility of the novel application of TPPs to temperature sensing for the first time. Based on a photonic extension of the Su-Schrieffer-Heeger (SSH) model \cite{suPRL1979}, schematically shown in Fig.\ref{schematic_chain}, we realize TPPs in 1D dimerized InSb microsphere chains in the terahertz range, protected by the band topology that is characterized by the quantized complex Zak phase. The choice of intrinsic InSb is motivated by its temperature-dependent carrier concentration that can result in a thermally-tunable plasma frequency in the terahertz range \cite{haleviPRL2000,gomezrivasPRB2006}. Based on this structure, we reveal that the resonance frequency of the TPPs can be successfully tuned by the temperature, leading to a high sensitivity and figure of merit. By calculating the local density of states (LDOS) at different temperatures, we further demonstrate that this temperature sensitivity can be experimentally detected and find that the LDOS signals of these TPPs are immune to disorder, leading to a robust temperature sensing functionality. 


In Fig.\ref{schematic_chain}, the chain is assumed to be aligned along the $x$-axis, where the dimerization is introduced by using inequivalent spacings $d_1$ and $d_2$ for the two sublattices, denoted by $A$ and $B$ with a dimerization parameter defined as $\beta=d_1/d$ where $d=d_1+d_2$ is the lattice constant.  This dimerization process gives rise to different “hopping” amplitudes of plasmon polaritons in either directions, well mimicking the SSH model for electrons \cite{suPRL1979}. Note due to the presence of near-field and far-field dipole-dipole interactions, a theoretical model beyond the nearest-neighbor approximation in the conventional SSH model should be implemented \cite{wang2018topological,wangPRB2018b}. To this end, the radius of the InSb microsphere is set to be $a=1\mathrm{\mu m}$, which is much smaller than the wavelength of interest (usually larger than $100\mathrm{\mu m}$). The electromagnetic response of an individual InSb microsphere is then described by the electric dipole polarizability with the radiative correction given by $\alpha(\omega)=\frac{4\pi a^3\alpha_0}{1-2i\alpha_0(ka)^3/3}$, where $\alpha_0(\omega)=\frac{\varepsilon_p(\omega)-1}{\varepsilon_p(\omega)+2}$, $\omega$ is the angular frequency of the driving field and $k=\omega/c$ is the wavenumber with $c$ denoting the speed of light in vacuum \cite{tervoPRMater2018,markelPRB2007,parkPRB2004}. The permittivity function of intrinsic InSb can be modeled by a Drude model as
$\varepsilon_p(\omega)=\varepsilon_\infty-\frac{\omega_p^2}{\omega^2+i\gamma\omega}$, where $\varepsilon_\infty=15.68$ is the high-frequency limit of the permittivity, $\omega_p$ is the plasma frequency, and $\gamma$ is the damping coefficient, both of which depend on the temperature \cite{cunninghamJAP1970,haleviPRL2000,hanJMO2009}. In particular,  $\omega_p=\sqrt{Ne^2/m^*\varepsilon_0}$ and $\gamma=e/m^*\mu$, where $N$ is the carrier concentration, $m^*$ is the effective mass of carries, $\mu$ is the mobility, $e$ is the electric charge of an electron and $\varepsilon_0$ is the permittivity of the vacuum. 

In the investigated temperature range in this work, the carrier mobility varies very slightly with the temperature and thus can be regarded as constant \cite{cunninghamJAP1970,howellsAPL1996}, leading to a decay rate of $\gamma=10\pi \times 10^{10}\mathrm{rad/s}$ \cite{haleviPRL2000,howellsAPL1996}. The effective mass is chosen to be $m^*=0.015m_e$ \cite{kittel1976}.  When the temperature $T$ is in the range from 160 to 350 K, the energy gap of InSb changes very little, and its the carrier concentration (in $\mathrm{cm}^{-3}$) as a function of temperature is described by the following experimental correlation equation  \cite{cunninghamJAP1970,haleviPRL2000,hanJMO2009}
\begin{equation}
N=5.76\times 10^{14}T^{\frac{3}{2}}\exp{(-\frac{0.13 \mathrm{eV}}{k_BT})},
\end{equation}
where $k_B$ is the Boltzmann constant. When the distance between the centers of different microspheres is larger than $3a$, the electromagnetic interactions are exactly described by the coupled-dipole equations \cite{parkPRB2004,markelPRB2007,tervoPRMater2018}:
\begin{equation}\label{coupled_dipole_eq}
\mathbf{p}_j(\omega)=\alpha(\omega)\left[\mathbf{E}_\mathrm{inc}(\mathbf{r}_j)+\frac{\omega^2}{c^2}\sum_{i=1,i\neq j}^{\infty}\mathbf{G}_0(\omega,\mathbf{r}_j,\mathbf{r}_i)\mathbf{p}_i(\omega)\right],
\end{equation}
where $\mathbf{E}_\mathrm{inc}(\mathbf{r})$ is the external incident field and $\mathbf{p}_j(\omega)$ is the excited electric dipole moment of the $j$-th microsphere. $\mathbf{G}_{0}(\omega,\mathbf{r}_j,\mathbf{r}_i)$ is the free-space dyadic Green's function describing the propagation of field emitting from the $i$-th microsphere to $j$-th microsphere \cite{markelPRB2007}. This model takes all types of near-field and far-field dipole-dipole interactions into account and is thus beyond the traditional nearest-neighbor approximation \cite{lingOE2015}. 

\begin{figure}[htbp]
	\centering
\subfloat{
	\includegraphics[width=0.46\linewidth]{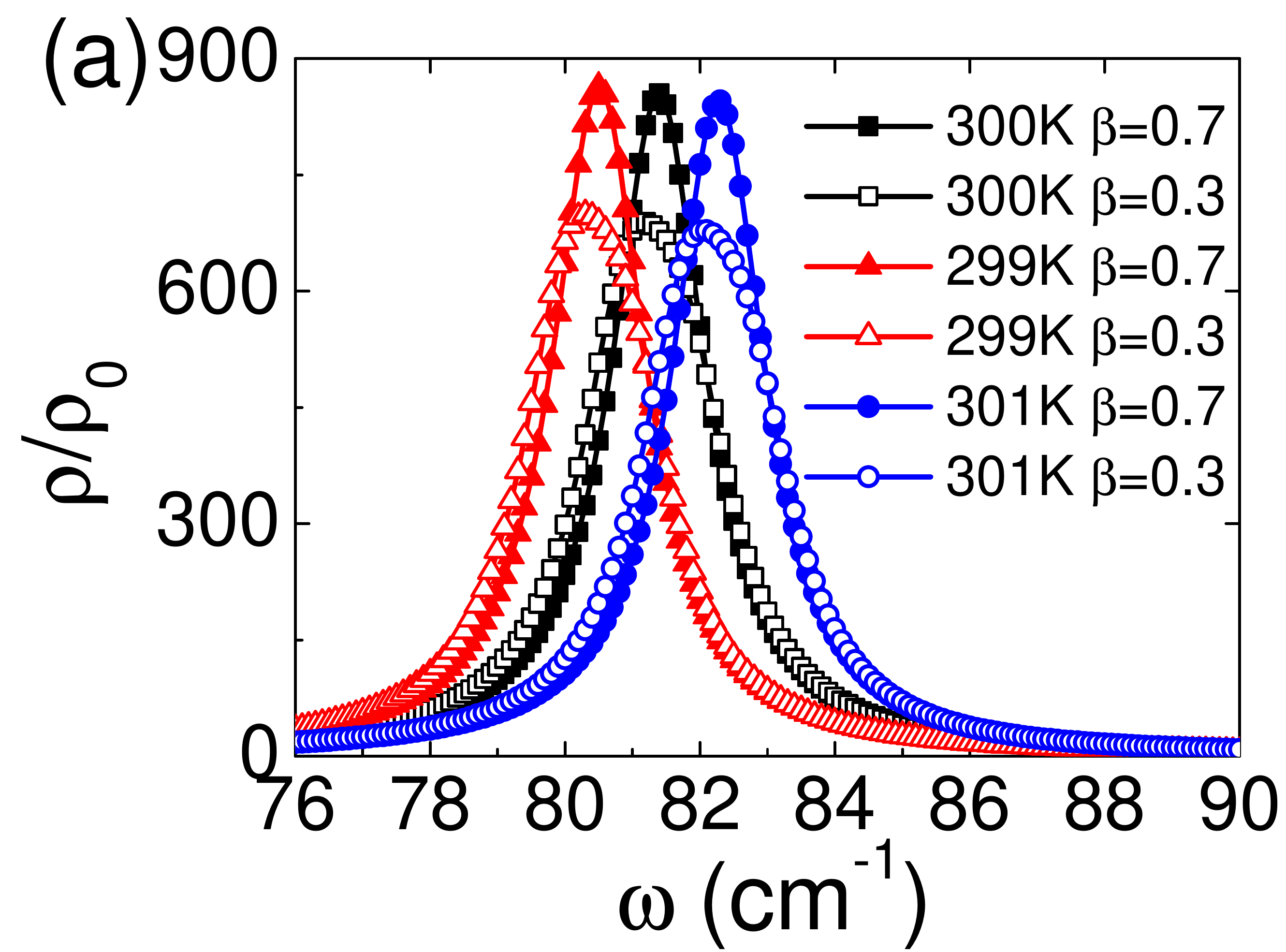}\label{LDOStemp}
}
\subfloat{
	\includegraphics[width=0.46\linewidth]{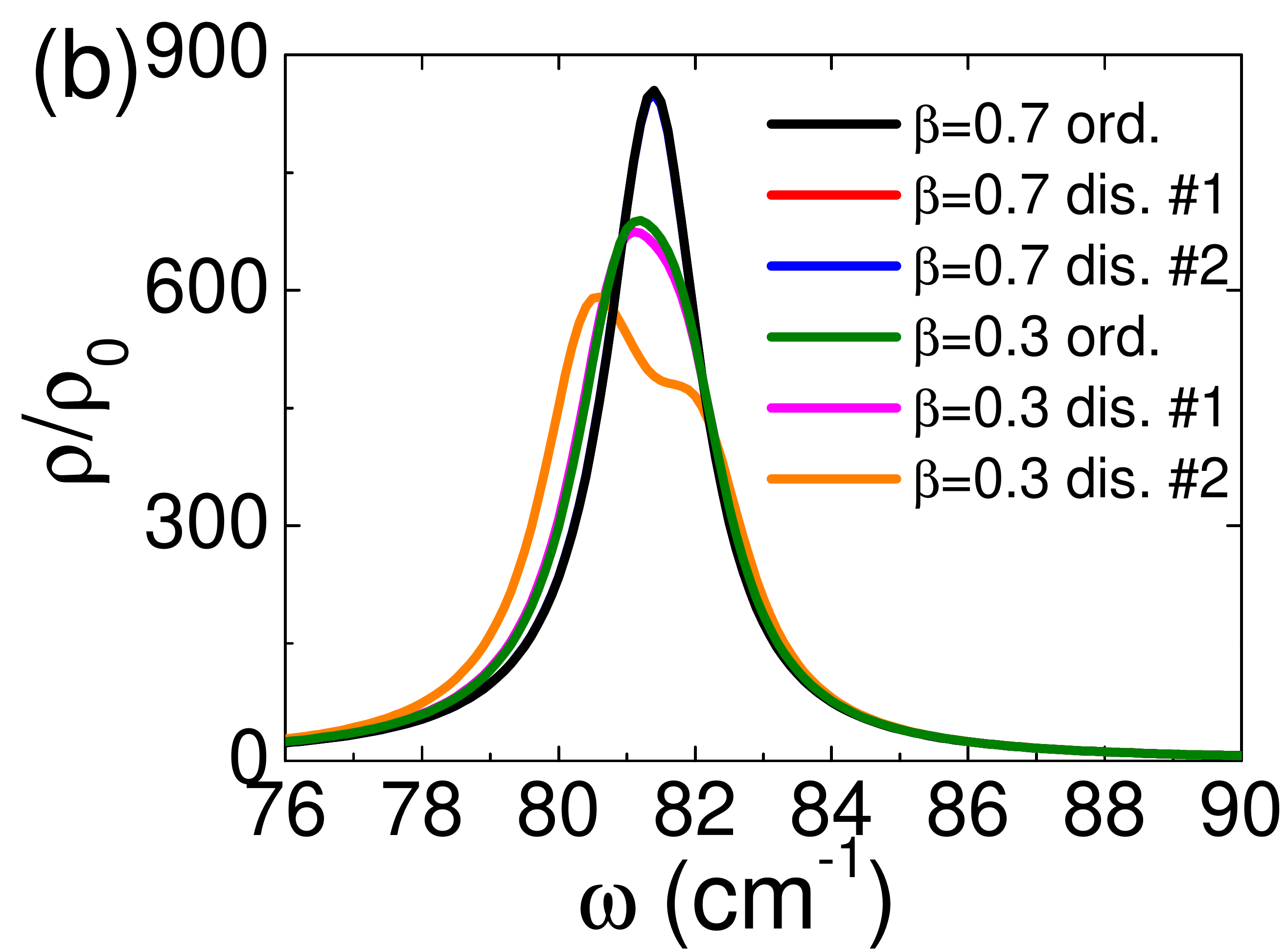}\label{LDOSdisorder}
}
	\caption{Calculated LDOS as an experimentally accessible signal of TPP temperature sensitivity. (a) The LDOS for topological and non-topological ordered chains at different temperatures. (b) The LDOS for topological and non-topological disordered (dis.) chains at $T$ = 300 K, compared with those of ordered (ord.) ones.}\label{LDOS}
	
\end{figure}

According to the polarization direction of the dipole moments, the eigenmodes can be divided into two types: transverse and longitudinal \cite{weberPRB2004}. More specifically, the dipole moments in the longitudinal modes are polarized along the $x$-axis, while those in the transverse modes are polarized perpendicular to the $x$-axis. In this work, we are mainly concerned with the topological properties of longitudinal modes. This is because transverse ones are more strongly coupled to the free-space radiation with a much narrower band gap and the localization degree is lower due to the long-range dipole-dipole interactions, all of which make it difficult to observe transverse topological eigenmodes experimentally, as discussed in our previous papers \cite{wang2018topological,wangPRB2018b}. We will also see that in Fig.\ref{LDOS}, the near-field detected signal is mainly determined by the longitudinal modes. In order to identify the topological properties, we first calculate the longitudinally polarized band structures under two different temperatures of $T=200~\mathrm{K}$ and 300 K, whose real parts are presented in Figs.\ref{bandstructurelongd10} and \ref{bandstructurelongd10T300} respectively for different dimerization parameters at a fixed lattice constant $d=10\mathrm{\mu m}$. The calculation is done by applying the Bloch theorem for an infinitely long chain \cite{wang2018topological,wangPRB2018b}. Since the band structure is identical for the cases of $\beta$ and $1-\beta$ with the difference lying in their topological invariant \cite{pocockArxiv2017,downingPRB2017}, the band structures for the cases of $\beta=0.3$ and $\beta=0.4$ are not plotted. For $\beta\neq0.5$, band gaps in the real frequency space persist to be open and a larger $|\beta-0.5|$ gives rise to a wider band gap. This behavior is consistent with the conventional SSH model \cite{atalaNaturephys2013}. Moreover, at different temperatures, the central frequency of the band gap is vastly different, actually close to the frequency of the localized surface plasmon resonance (LSPR) of a single microsphere at different temperatures, as a result of the strong coupling between the collective, delocalized surface plasmon polariton (SPP) modes of the two sublattices at $\beta\neq0.5$ \cite{wangPRB2018b}.

While the present system is open and hence non-Hermitian, for longitudinal modes the non-Hermiticity in essence does not break the bulk-boundary correspondence and the complex Zak phase $\theta_\mathrm{Z}$ is adequate to capture the topological properties in the bulk side, which is the geometric phase picked up by an eigenmode when it adiabatically evolves across the first Brillouin zone (BZ), as proved by previous works \cite{lieuPRB2018,wangPRB2018b,wang2018topological}, from which more details can be found. Moreover, although in the present system the chiral symmetry breaks down, $\theta_\mathrm{Z}$ is still quantized (having only two values, 0 and $\pi$) in a similar way as in the chirally symmetric system because the eigenvectors are independent of the chiral-symmetry breaking terms in the effective Hamiltonian \cite{pocockArxiv2017,wang2018topological,wangPRB2018b}. Calculation shows that the complex Zak phase for $\beta=0.7$ and $\beta=0.6$ is $\pi$ and $0$ for $\beta=0.3$ and $\beta=0.4$. Furthermore, we have examined that regardless of the lattice constant and temperature, the complex Zak phase is guaranteed to be 0 for $\beta<0.5$ and $\pi$ for $\beta>0.5$ for longitudinal modes \cite{wang2018topological,wangPRB2018b}. Consequently in the present system, the complex Zak phase is a well-defined topological invariant in the bulk side.

\begin{figure}[htbp]
	\centering
	\subfloat{
	\includegraphics[width=0.47\linewidth]{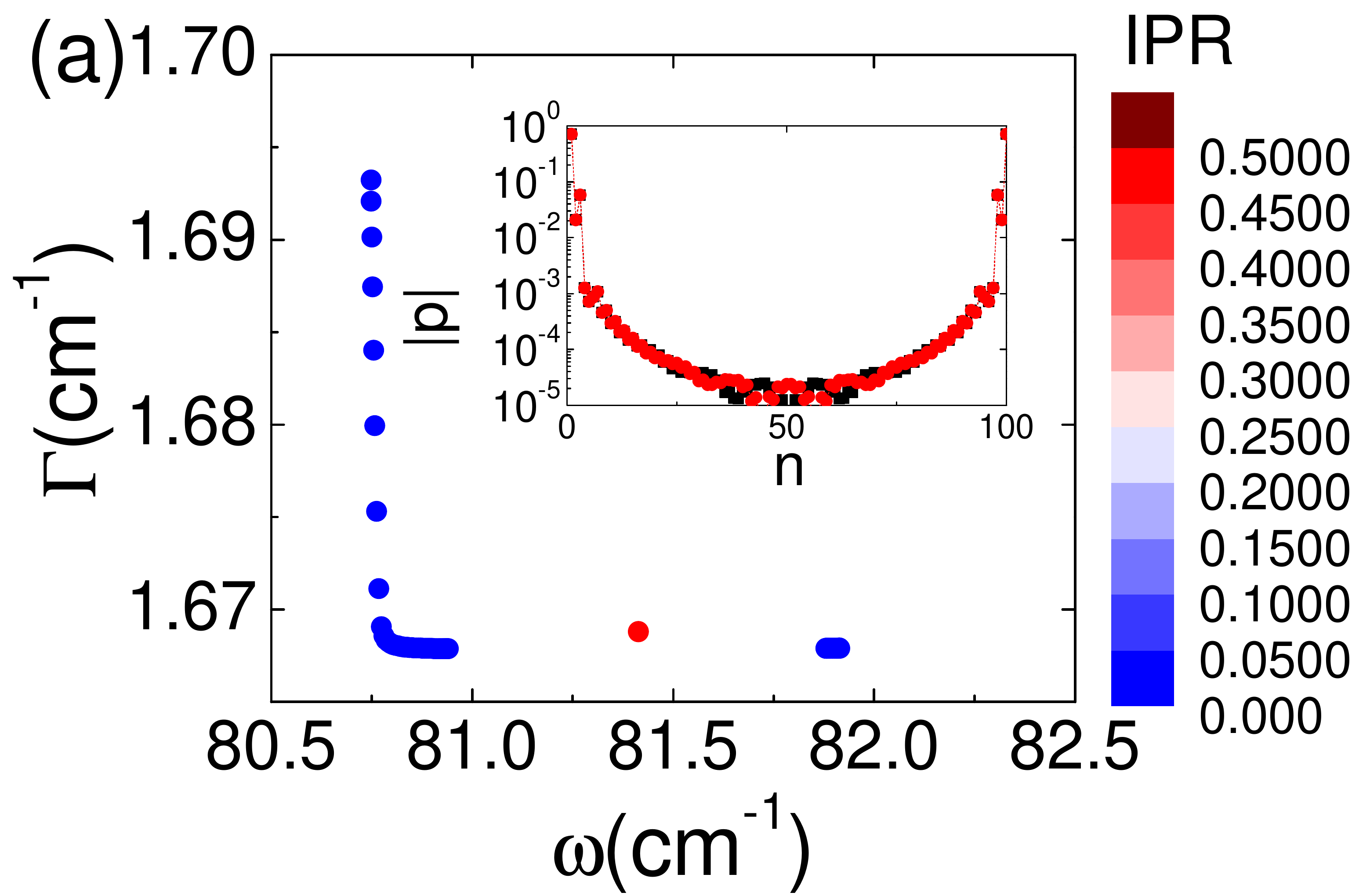}\label{beta07longband}
}
	\hspace{0.01in}
	\subfloat{
	\includegraphics[width=0.45\linewidth]{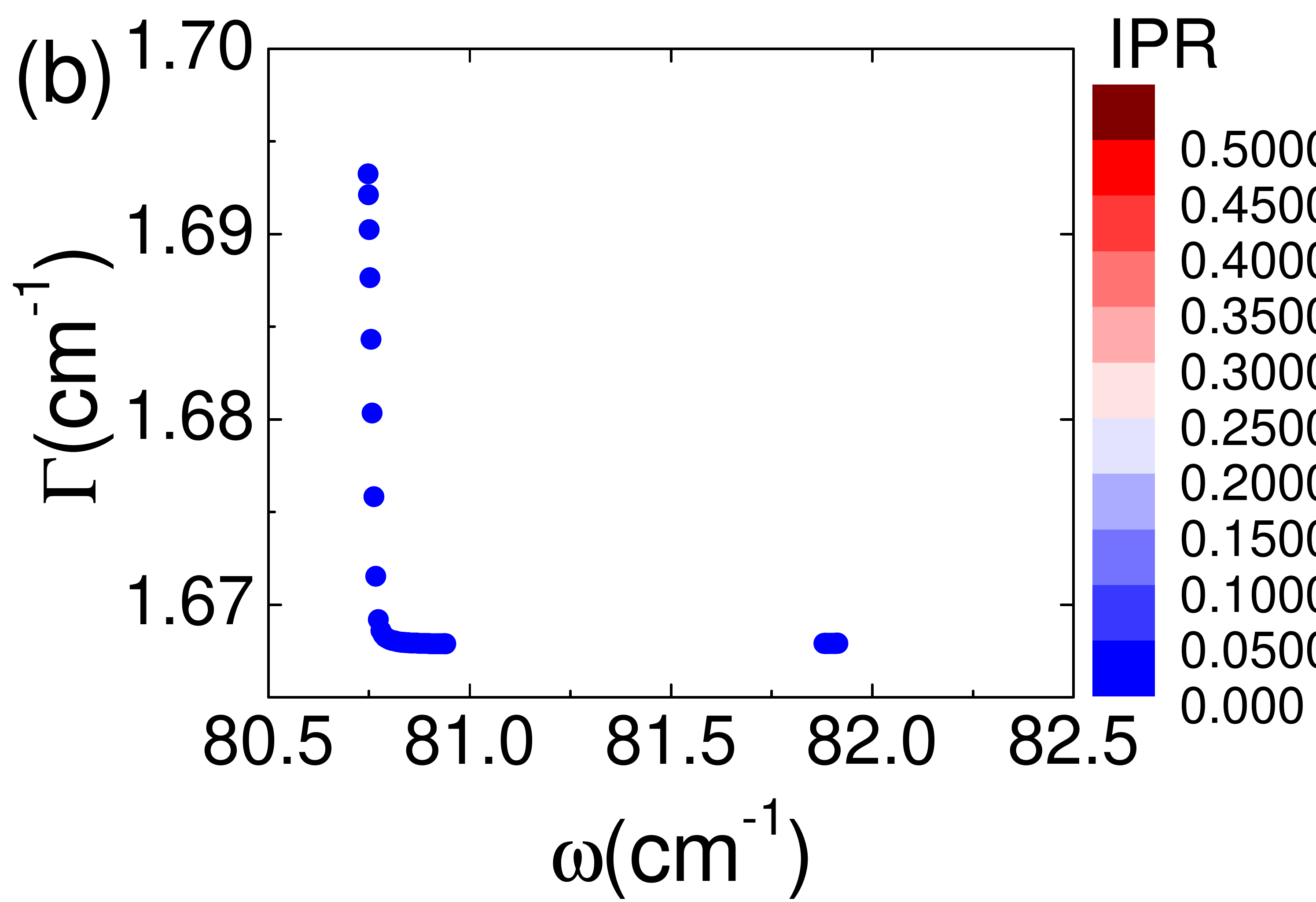}\label{beta03longband}
}
	\hspace{0.01in}
	\subfloat{
	\includegraphics[width=0.5\linewidth]{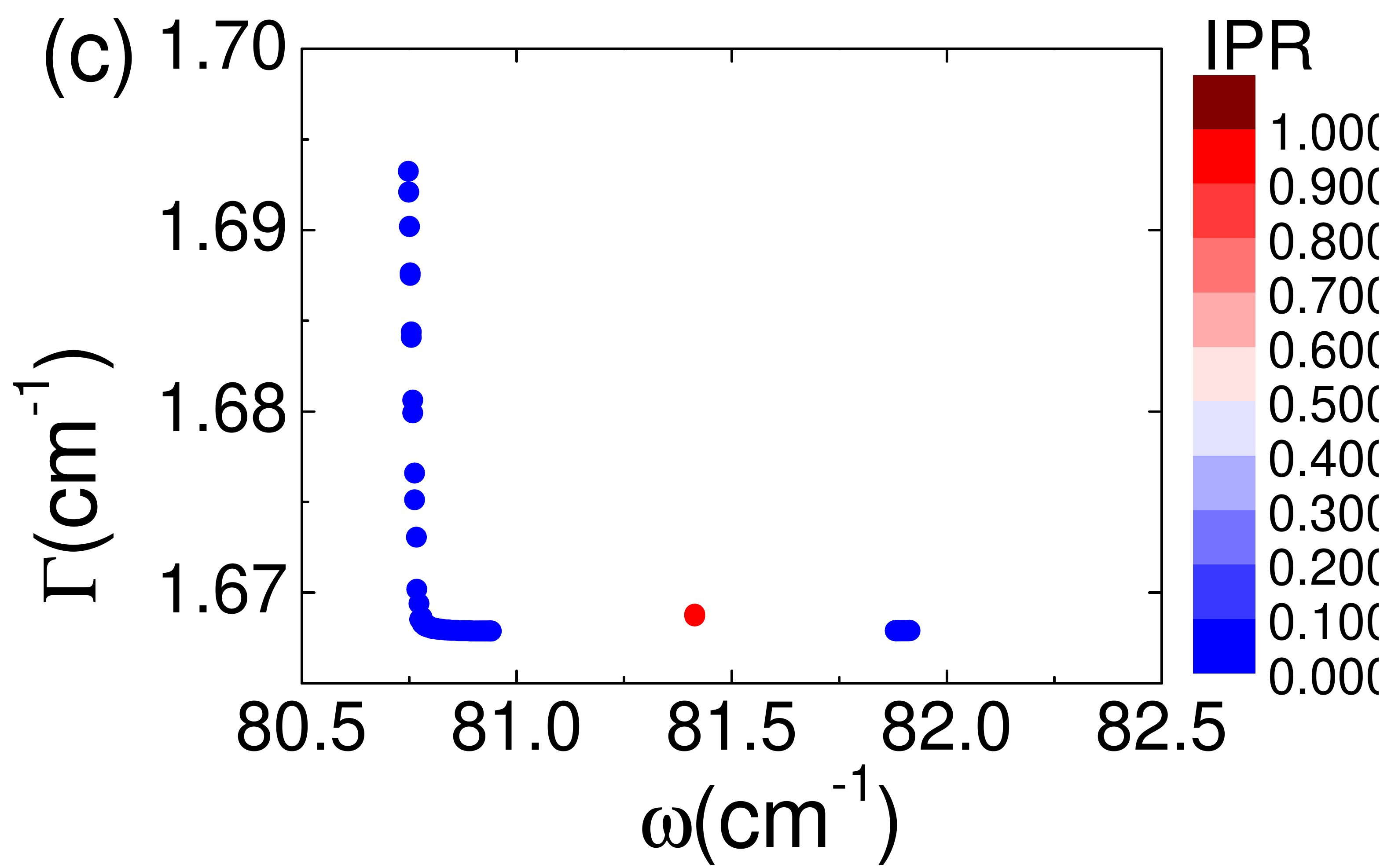}\label{interfacemodebeta07}
}
\hspace{0.01in}
\subfloat{
	\includegraphics[width=0.43\linewidth]{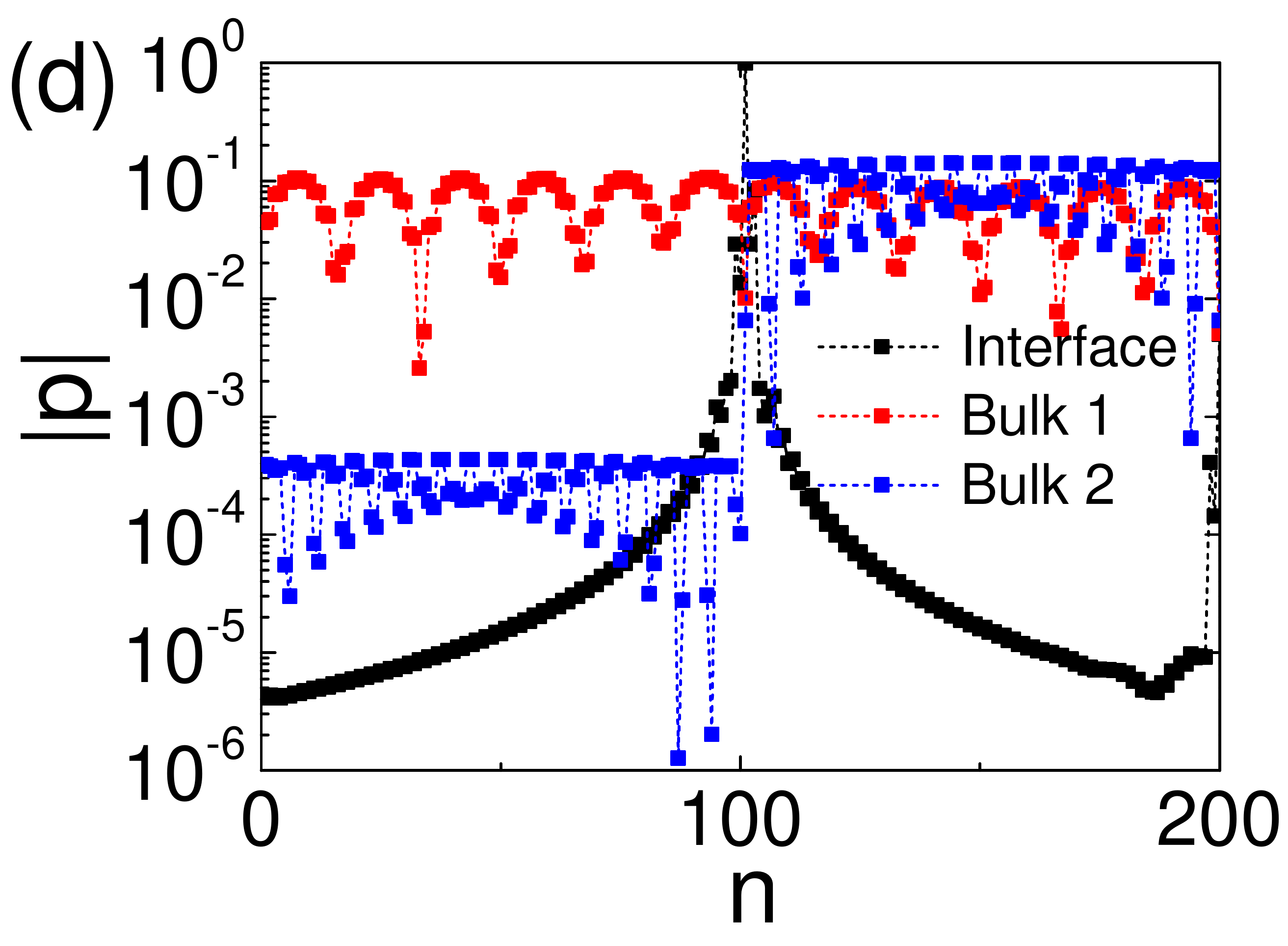}\label{interfacemodedipole}
}
\hspace{0.01in}
	\subfloat{
	\includegraphics[width=0.51\linewidth]{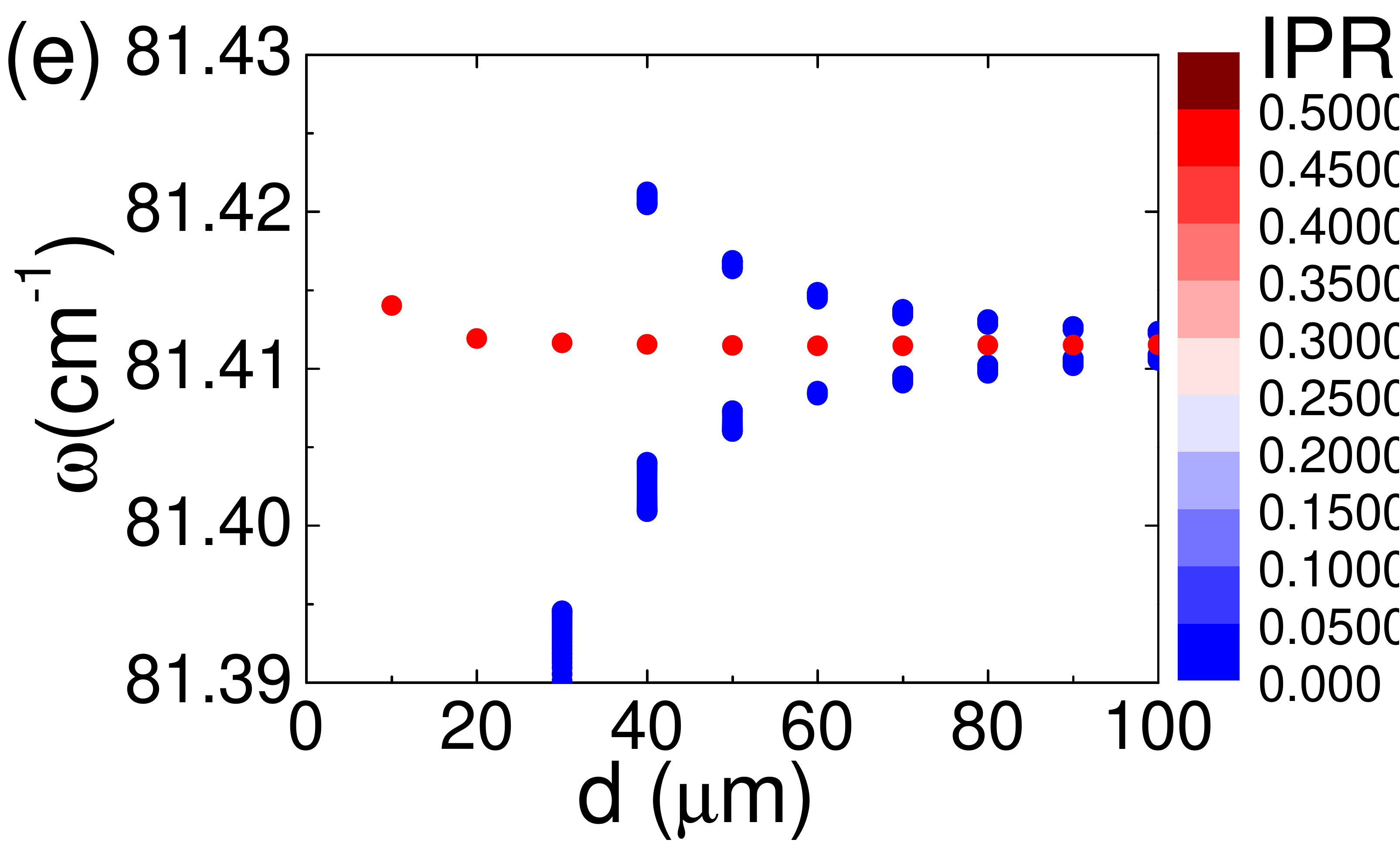}\label{bandevolutionlongreal}
}
\hspace{0.01in}
\subfloat{
	\includegraphics[width=0.42\linewidth]{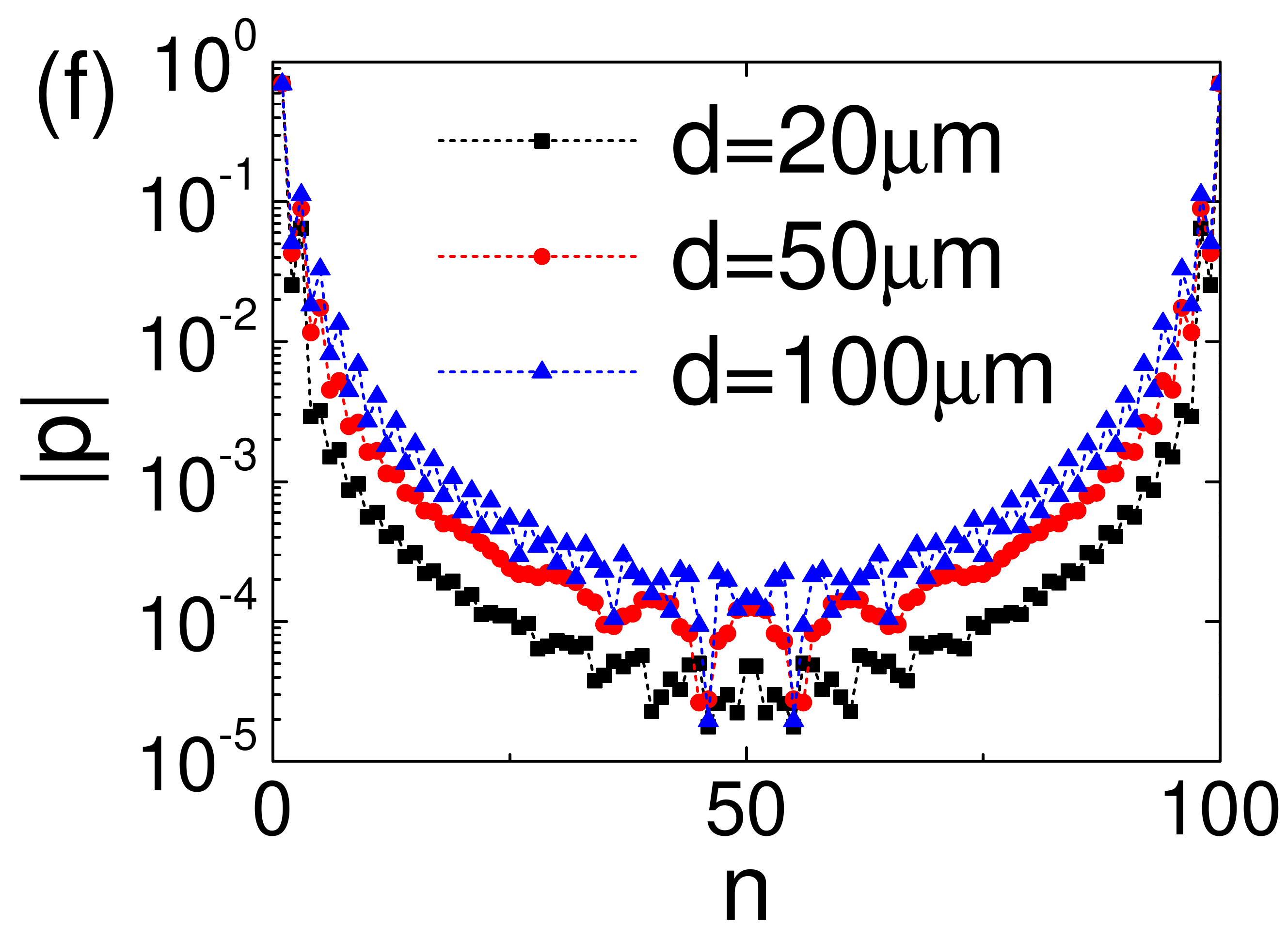}\label{midgapmodewithd}
}

	\caption{Topological eigenmodes in finite systems at $T=300~\mathrm{K}$. (a) Longitudinal eigenmode distribution of a dimerized chain with $N=100$ microspheres under $\beta=0.7$ and $d=10\mathrm{\mu m}$. Note there are two spectrally overlapping midgap modes. Inset: Dipole moment distribution of the midgap edge modes. (b) The same as (a) but here $\beta=0.3$. (c) Longitudinal eigenmode distribution for a connected chain. (d) Dipole moment distribution of the interface mode in (c), compared with those of two arbitrarily chosen bulk eigenmodes. (e) Real parts of the complex eigenfrequency spectrum of longitudinal eigenmodes as a function of the lattice constant $d$ for $\beta=0.7$. (f) Dipole moment distribution for the topological edge modes at different lattice constants.} \label{eigenmodelong}
\end{figure}

To examine the bulk-boundary correspondence, a crucial principle in topological physics, we then turn to a finite system under the open boundary condition and calculate its eigenmode distribution (i.e., discrete band structures). This can be done by using Eq.(\ref{coupled_dipole_eq}) with a zero incident field \cite{weberPRB2004,pocockArxiv2017}. More specifically, an eigenvalue equation in the form of $\mathbf{G}|\mathbf{p}\rangle=\alpha^{-1}(\omega)|\mathbf{p}\rangle$ is obtained, with $\mathbf{G}$ standing for the interaction Green's matrix and $|\mathbf{p}\rangle=[p_1p_2...p_j...p_N]$ denoting the right eigenvector or the dipole moment distribution of an eigenmode with $p_j$ the dipole moment of the $j$-th microsphere. This equation gives rises to a set of complex eigenfrequencies in the lower complex plane like $\tilde{\omega}=\omega-i\Gamma/2$, where the real part $\omega$ stands for the angular frequency while the imaginary part $\Gamma$ corresponds to the linewidth (or decay rate) of the eigenmode \cite{pocockArxiv2017,wang2018topological,wangPRB2018b}. Additionally, for each eigenmode, the inverse participation ratio (IPR) is calculated $\mathrm{IPR}=\frac{\sum_{n=1}^{N}|p_j|^4}{(\sum_{n=1}^{N}|p_j|^2)^2}$, which is closer to 1 for a more spatially localized eigenmode \cite{wangOL2018,wang2018topological}. The results are presented in Figs.\ref{beta07longband} and \ref{beta03longband} for the cases of $\beta=0.7$ and $\beta=0.3$ under $T=300~\mathrm{K}$. In both cases complex band gaps keep open, consistent with the complex band gaps in the Bloch band structure. However, a significant difference between the $\beta=0.7$ and $\beta=0.3$ cases can be noted, which is that there are two midgap modes with high IPRs in the band gap in the former case. The dipole moment distributions of the two midgap modes are shown in the inset of Fig.\ref{beta07longband}, which is found to be highly localized over both boundaries. By considering the nontrivial complex Zak phase of the $\beta=0.7$ case in the bulk side, it can be verified that these midgap modes are topologically protected edge modes, namely, topological plasmon polaritons.

%
%

To further demonstrate the bulk-boundary correspondence, Fig.\ref{interfacemodebeta07} shows the eigenmode distribution of a 1D connected chain consisting of a topologically trivial chain with $\beta=0.3$ in the left and a topologically nontrivial chain with $\beta=0.7$ in the right. The distance between the two chains is set to be $10\mathrm{\mu m}$. We can clearly observe two midgap modes with high IPRs reaching over 0.9, one of which is the topological interface mode while the other is the topological edge mode localized at the right boundary of the right chain. In addition, in Fig.\ref{interfacemodedipole}, the dipole moment distribution for the topological interface mode is shown compared to those of two typical bulk eigenmodes, which are extended over the chain. To investigate the effect of lattice constant on the topological edge modes, the real part of the eigenfrequency spectrum of a finite chain as a function of the lattice constant are presented in Fig.\ref{bandevolutionlongreal} with the dimerization parameter fixed as $\beta=0.7$. It is found that the complex band gaps persist to be open with high-IPR eigenmodes robustly emerging in the complex band gaps. Note at large lattice constants, the real band gap almost closes while the imaginary part still opens, which are not shown here for brevity.  In addition, the complex frequency of the topological modes hardly varies with the increase of the lattice constant, indicating its robustness (with a variation of angular frequency smaller than 0.003 cm$^{-1}$ or about 5 nm). The dipole moment distributions of the topological midgap modes under different lattice constants are also presented in Fig.\ref{midgapmodewithd}. In addition, we also confirm that for the cases of $\beta<0.5$ at different lattice constants, no localized edge eigenmodes can be found. Therefore, by summarizing the results presented in Figs.\ref{figcda} and \ref{eigenmodelong}, we can unambiguously confirmed that highly localized TPPs are found in the present system, which are topologically protected by the well-defined complex Zak phase if the dimerization parameter $\beta>0.5$, regardless of the lattice constant and temperature.

\begin{figure}[htbp]
	\centering
	\subfloat{
		\includegraphics[width=0.41\linewidth]{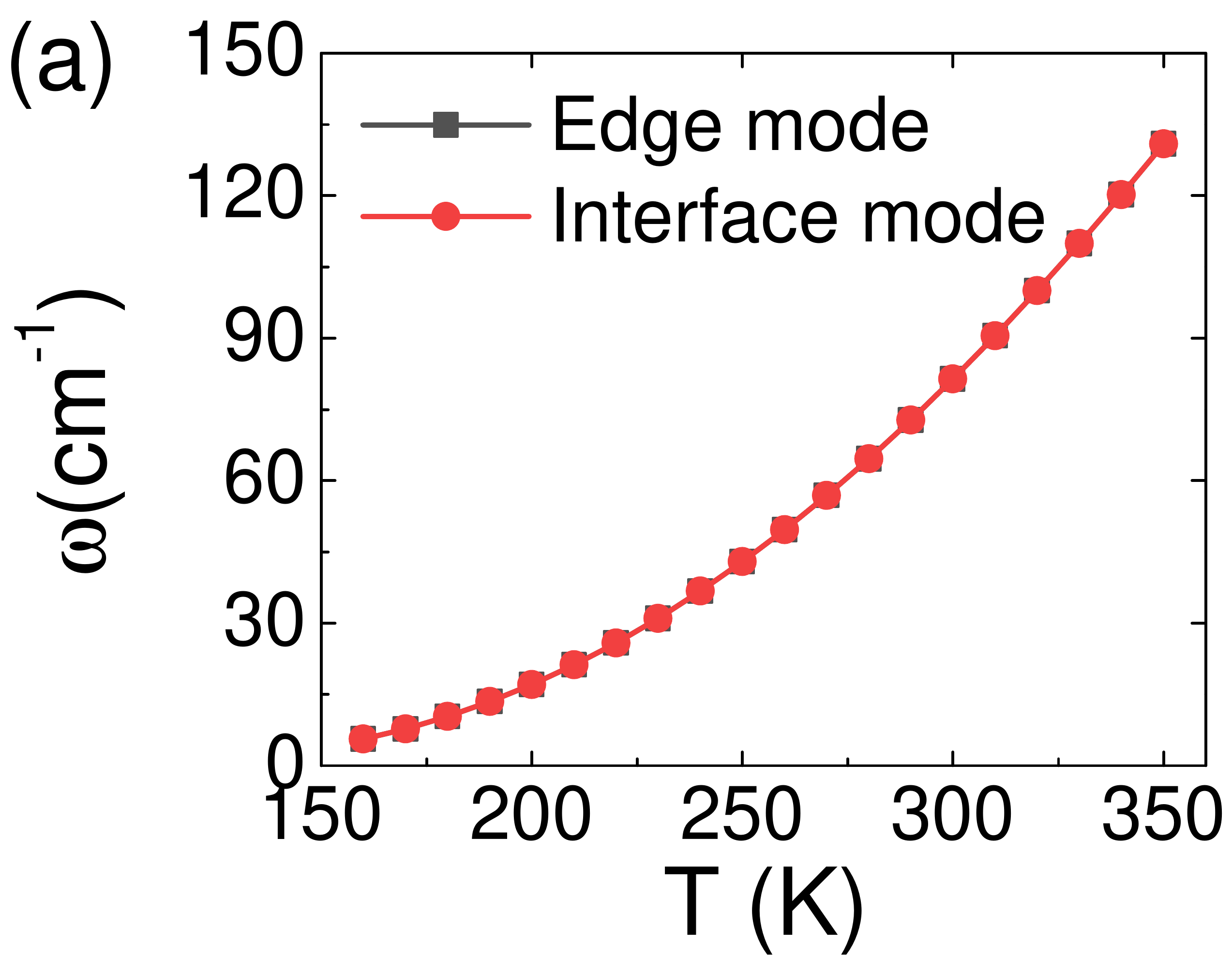}\label{tempeffect}
	}
    \hspace{0.01in}
\subfloat{
	\includegraphics[width=0.51\linewidth]{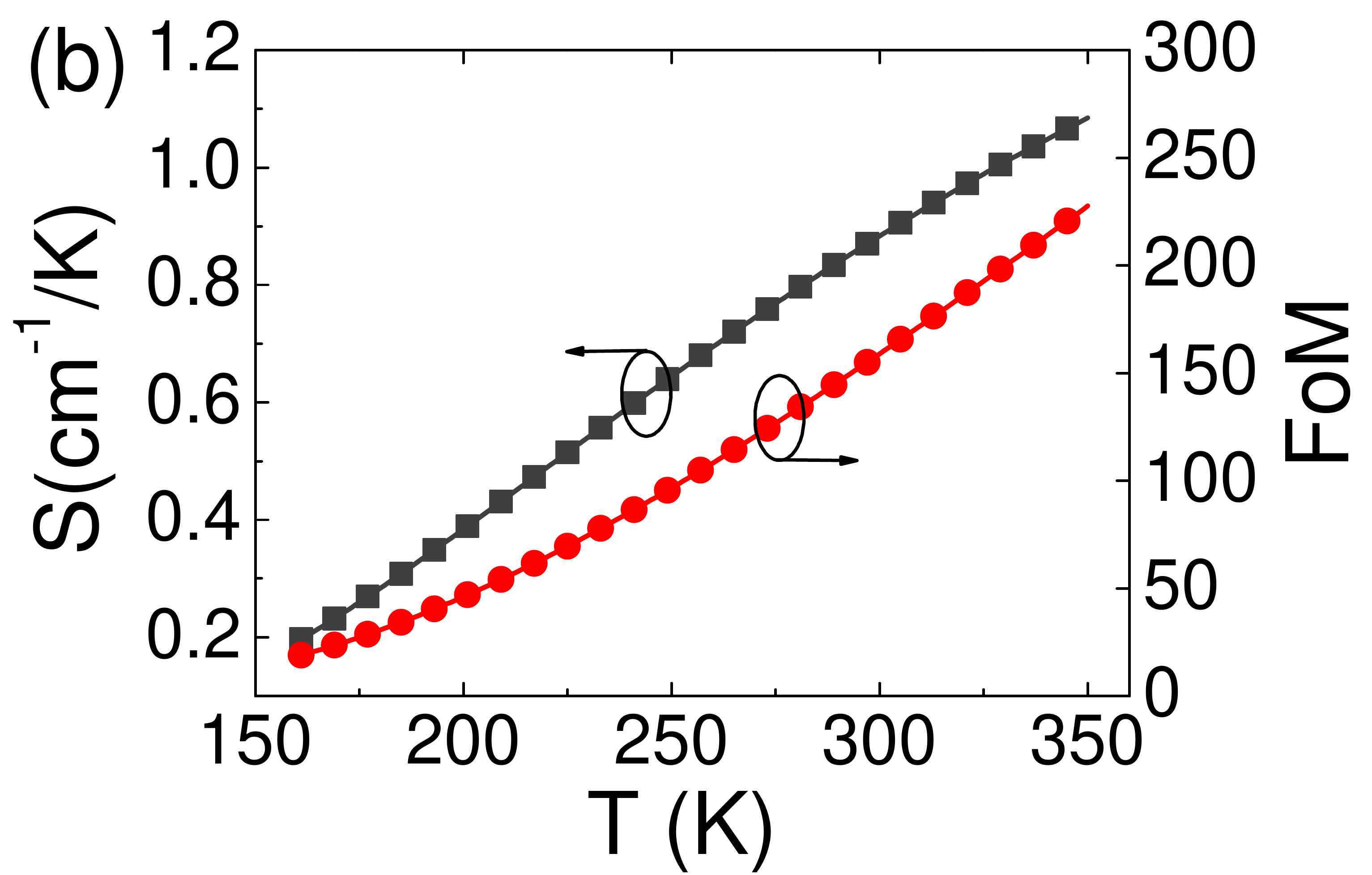}\label{sensitivity}
}

\caption{(a) Effects of temperature on  the resonance frequency (wavelength) of edge and interface modes of the topological plasmon polaritons. (b) Sensitivity and figure of merit (FoM) quantifying the temperature sensing performance of the topological plasmon polaritons. Here the parameters are $\beta=0.7$ and $d=10 \mathrm{\mu m}$.}\label{temp1}
	
\end{figure}

On the basis of above analysis, we continue to investigate the effects of temperature on these TPPs and their application to temperature sensing. In Fig.\ref{temp1}, we show the variation of resonance frequencies of edge and interface TPPs with the temperature. It is observed that the resonance frequency of TPPs can be varied in a wide frequency range from about 0.2 THz to 4 THz by increasing the temperature from 160 K to 350 K. The frequencies of edge and interface TPPs almost overlap with each other, indicating the spectral stability of these topological modes against geometric parameters. The band gaps also become wider as the temperature grows, which makes it easier for experimental observation of TPPs at relatively high temperatures (not shown here). These observations feature a temperature sensing functionality for the TPPs. The temperature sensitivity can be defined from the resonance frequency of TPPs as $S=\Delta \omega_\mathrm{TPP}/\Delta T$, where $\Delta$ indicates a small variation for a physical quantity. At room temperature (approximately taken 300 K), the sensitivity is about $0.88~ \mathrm{cm}^{-1}/\mathrm{K}$, or $0.0264~\mathrm{THz/K}$, which is equivalent a wavelength shift around $1.3423~\mathrm{\mu m/K}$. Moreover, we can also define a figure of merit (FoM) of this temperature sensing performance as $\mathrm{FoM}=ST/\Gamma=\Delta \omega_\mathrm{TPP}T/(\Gamma\Delta T)$, which reaches over 150 at room temperature. In Fig.\ref{sensitivity}, the variation of $S$ and FoM with the temperature are presented. As a comparison, recent optical temperature sensor has a sensitivity about $8.9\times 10^{-3}~\mathrm{THz/K}$ and FoM around 119 \cite{maIEEEJSTQE2017}, which are significantly lower.

In experiment, it has been shown that TPPs, or more generally, topological photonic modes, can be conveniently detected by the scanning near-field optical microscopy (SNOM) \cite{proctor2019exciting,slobozhanyukAPL2019} or other high-momentum sources \cite{pengPRL2019}. Here we numerically calculate LDOS $\rho$ excited by an ideal electric dipole near the chain with a distance of 100 nm at different temperatures as a proof-of-concept demonstration, which is also schematically shown in Fig.\ref{schematic_chain} with a SNOM cantilever nearby. The calculation details can be found in Ref.\cite{wangPRB2018b}. The results of normalized LDOS $\rho/\rho_0$ for both topologically nontrivial and trivial systems are summarized in Fig.\ref{LDOStemp}, where $\rho_0=\omega^2/(\pi^2c^3)$ is the LDOS in vacuum. It is found that for topological chains ($\beta=0.7$), LDOS, as a manifestation of light-matter interactions, is substantially enhanced in the band gaps due to the existence of TPPs, while for topologically trivial chains ($\beta=0.3$), the LDOS peaks emerge at the band edges. Note carefully there is a slight shift between the LDOS peaks of topological trivial and nontrivial cases \cite{wangPRB2018b}. For example, for $T=300~\mathrm{K}$ at the spectral position of TPPs, namely, $\omega=81.4~\mathrm{cm}^{-1}$, $\rho/\rho_0=854.87$, which reaches its maximum for the topological chain, while at this angular frequency $\rho/\rho_0=676.86$ for the non-topological chain and the peak position is at around $\omega=81.2~\mathrm{cm}^{-1}$. Moreover, by varying the temperature to 299 K and 301 K, we can observe considerable shifts of LDOS peaks. Therefore, the temperature sensitivity of TPPs is experimentally detectable via near-field probing techniques like SNOM and terahertz quantum emitters \cite{wuACSANM2019}. 

Furthermore, we can confirm that the enhancement and spectral positions of LDOS due to TPPs are robust over disorder. The disorder is introduced by shifting the positions of $B$-type NPs randomly in the range $[-\eta d_1/2,\eta d_1/2]$ along the $x$-axis, while the positions of $A$-type NPs are fixed, to make period the $d$ constant. Such disorder breaks both inversion and chiral symmetries, at least one of which are regarded as protecting the topological properties of the present system \cite{wangPRB2018b,lieuPRB2018,pocockArxiv2017,pocockNanophoton2019}. In Fig.\ref{LDOSdisorder}, results of the LDOS at $T$ = 300 K of two randomly chosen disordered realizations for both topological and non-topological chains are presented, compared with those of ordered ones. It is clearly seen that the LDOS peaks in topological chains are very robust against disorder while those of non-topological chains significantly vary with the detailed positions of microspheres. 

To summarize, we theoretically investigate the application of TPPs to temperature sensing for the first time. We show that TPPs can be realized in a one-dimensional intrinsic InSb microsphere chain. By utilizing the temperature dependence of the permittivity of InSb, the resonance frequency of the TPPs can be thermally tuned. Moreover, the temperature sensitivity of the TPPs can be as high as $0.0264~\mathrm{THz/K}$ at room temperature, leading to a FoM over 150.  By calculating the LDOS near the chain as an experimentally detectable signal, we further demonstrate that these TPPs can achieve a strong confinement of radiation and are also immune to disorder. We envisage these TPPs can be utilized as promising candidates for robust and enhanced temperature sensing, especially for on-chip thermometry \cite{nguyenAPL2018}.  

\begin{acknowledgments}
This work is  supported by the National Natural Science Foundation of China (No. 51636004 and No. 51906144), Shanghai Key Fundamental Research Grant (No. 18JC1413300), China Postdoctoral Science Foundation (No. BX20180187 and No. 2019M651493) and the Foundation for Innovative Research Groups of the National Natural Science Foundation of China (No. 51521004).
\end{acknowledgments}
\bibliography{ssh_insb}

\begin{thebibliography}{42}%
\makeatletter
\providecommand \@ifxundefined [1]{%
 \@ifx{#1\undefined}
}%
\providecommand \@ifnum [1]{%
 \ifnum #1\expandafter \@firstoftwo
 \else \expandafter \@secondoftwo
 \fi
}%
\providecommand \@ifx [1]{%
 \ifx #1\expandafter \@firstoftwo
 \else \expandafter \@secondoftwo
 \fi
}%
\providecommand \natexlab [1]{#1}%
\providecommand \enquote  [1]{``#1''}%
\providecommand \bibnamefont  [1]{#1}%
\providecommand \bibfnamefont [1]{#1}%
\providecommand \citenamefont [1]{#1}%
\providecommand \href@noop [0]{\@secondoftwo}%
\providecommand \href [0]{\begingroup \@sanitize@url \@href}%
\providecommand \@href[1]{\@@startlink{#1}\@@href}%
\providecommand \@@href[1]{\endgroup#1\@@endlink}%
\providecommand \@sanitize@url [0]{\catcode `\\12\catcode `\$12\catcode
  `\&12\catcode `\#12\catcode `\^12\catcode `\_12\catcode `\%12\relax}%
\providecommand \@@startlink[1]{}%
\providecommand \@@endlink[0]{}%
\providecommand \url  [0]{\begingroup\@sanitize@url \@url }%
\providecommand \@url [1]{\endgroup\@href {#1}{\urlprefix }}%
\providecommand \urlprefix  [0]{URL }%
\providecommand \Eprint [0]{\href }%
\providecommand \doibase [0]{http://dx.doi.org/}%
\providecommand \selectlanguage [0]{\@gobble}%
\providecommand \bibinfo  [0]{\@secondoftwo}%
\providecommand \bibfield  [0]{\@secondoftwo}%
\providecommand \translation [1]{[#1]}%
\providecommand \BibitemOpen [0]{}%
\providecommand \bibitemStop [0]{}%
\providecommand \bibitemNoStop [0]{.\EOS\space}%
\providecommand \EOS [0]{\spacefactor3000\relax}%
\providecommand \BibitemShut  [1]{\csname bibitem#1\endcsname}%
\let\auto@bib@innerbib\@empty
\bibitem [{\citenamefont {Hasan}\ and\ \citenamefont
  {Kane}(2010)}]{hasanRMP2010}%
  \BibitemOpen
  \bibfield  {author} {\bibinfo {author} {\bibfnamefont {M.~Z.}\ \bibnamefont
  {Hasan}}\ and\ \bibinfo {author} {\bibfnamefont {C.~L.}\ \bibnamefont
  {Kane}},\ }\href {\doibase 10.1103/RevModPhys.82.3045} {\bibfield  {journal}
  {\bibinfo  {journal} {Rev. Mod. Phys.}\ }\textbf {\bibinfo {volume} {82}},\
  \bibinfo {pages} {3045} (\bibinfo {year} {2010})}\BibitemShut {NoStop}%
\bibitem [{\citenamefont {Ozawa}\ \emph {et~al.}(2019)\citenamefont {Ozawa},
  \citenamefont {Price}, \citenamefont {Amo}, \citenamefont {Goldman},
  \citenamefont {Hafezi}, \citenamefont {Lu}, \citenamefont {Rechtsman},
  \citenamefont {Schuster}, \citenamefont {Simon}, \citenamefont {Zilberberg},\
  and\ \citenamefont {Carusotto}}]{ozawa2018topological}%
  \BibitemOpen
  \bibfield  {author} {\bibinfo {author} {\bibfnamefont {T.}~\bibnamefont
  {Ozawa}}, \bibinfo {author} {\bibfnamefont {H.~M.}\ \bibnamefont {Price}},
  \bibinfo {author} {\bibfnamefont {A.}~\bibnamefont {Amo}}, \bibinfo {author}
  {\bibfnamefont {N.}~\bibnamefont {Goldman}}, \bibinfo {author} {\bibfnamefont
  {M.}~\bibnamefont {Hafezi}}, \bibinfo {author} {\bibfnamefont
  {L.}~\bibnamefont {Lu}}, \bibinfo {author} {\bibfnamefont {M.~C.}\
  \bibnamefont {Rechtsman}}, \bibinfo {author} {\bibfnamefont {D.}~\bibnamefont
  {Schuster}}, \bibinfo {author} {\bibfnamefont {J.}~\bibnamefont {Simon}},
  \bibinfo {author} {\bibfnamefont {O.}~\bibnamefont {Zilberberg}}, \ and\
  \bibinfo {author} {\bibfnamefont {I.}~\bibnamefont {Carusotto}},\ }\href
  {\doibase 10.1103/RevModPhys.91.015006} {\bibfield  {journal} {\bibinfo
  {journal} {Rev. Mod. Phys.}\ }\textbf {\bibinfo {volume} {91}},\ \bibinfo
  {pages} {015006} (\bibinfo {year} {2019})}\BibitemShut {NoStop}%
\bibitem [{\citenamefont {He}\ \emph {et~al.}(2016)\citenamefont {He},
  \citenamefont {Ni}, \citenamefont {Ge}, \citenamefont {Sun}, \citenamefont
  {Chen}, \citenamefont {Lu}, \citenamefont {Liu},\ and\ \citenamefont
  {Chen}}]{heNaturephys2016}%
  \BibitemOpen
  \bibfield  {author} {\bibinfo {author} {\bibfnamefont {C.}~\bibnamefont
  {He}}, \bibinfo {author} {\bibfnamefont {X.}~\bibnamefont {Ni}}, \bibinfo
  {author} {\bibfnamefont {H.}~\bibnamefont {Ge}}, \bibinfo {author}
  {\bibfnamefont {X.-C.}\ \bibnamefont {Sun}}, \bibinfo {author} {\bibfnamefont
  {Y.-B.}\ \bibnamefont {Chen}}, \bibinfo {author} {\bibfnamefont {M.-H.}\
  \bibnamefont {Lu}}, \bibinfo {author} {\bibfnamefont {X.-P.}\ \bibnamefont
  {Liu}}, \ and\ \bibinfo {author} {\bibfnamefont {Y.-F.}\ \bibnamefont
  {Chen}},\ }\href {https://doi.org/10.1038/nphys3867} {\bibfield  {journal}
  {\bibinfo  {journal} {Nature Physics}\ }\textbf {\bibinfo {volume} {12}},\
  \bibinfo {pages} {1124 EP } (\bibinfo {year} {2016})}\BibitemShut {NoStop}%
\bibitem [{\citenamefont {Atala}\ \emph {et~al.}(2013)\citenamefont {Atala},
  \citenamefont {Aidelsburger}, \citenamefont {Barreiro}, \citenamefont
  {Abanin}, \citenamefont {Kitagawa}, \citenamefont {Demler},\ and\
  \citenamefont {Bloch}}]{atalaNaturephys2013}%
  \BibitemOpen
  \bibfield  {author} {\bibinfo {author} {\bibfnamefont {M.}~\bibnamefont
  {Atala}}, \bibinfo {author} {\bibfnamefont {M.}~\bibnamefont {Aidelsburger}},
  \bibinfo {author} {\bibfnamefont {J.~T.}\ \bibnamefont {Barreiro}}, \bibinfo
  {author} {\bibfnamefont {D.}~\bibnamefont {Abanin}}, \bibinfo {author}
  {\bibfnamefont {T.}~\bibnamefont {Kitagawa}}, \bibinfo {author}
  {\bibfnamefont {E.}~\bibnamefont {Demler}}, \ and\ \bibinfo {author}
  {\bibfnamefont {I.}~\bibnamefont {Bloch}},\ }\href@noop {} {\bibfield
  {journal} {\bibinfo  {journal} {Nature Physics}\ }\textbf {\bibinfo {volume}
  {9}},\ \bibinfo {pages} {795} (\bibinfo {year} {2013})}\BibitemShut {NoStop}%
\bibitem [{\citenamefont {S{\"u}sstrunk}\ and\ \citenamefont
  {Huber}(2015)}]{susstrunkScience2015}%
  \BibitemOpen
  \bibfield  {author} {\bibinfo {author} {\bibfnamefont {R.}~\bibnamefont
  {S{\"u}sstrunk}}\ and\ \bibinfo {author} {\bibfnamefont {S.~D.}\ \bibnamefont
  {Huber}},\ }\href {\doibase 10.1126/science.aab0239} {\bibfield  {journal}
  {\bibinfo  {journal} {Science}\ }\textbf {\bibinfo {volume} {349}},\ \bibinfo
  {pages} {47} (\bibinfo {year} {2015})},\ \Eprint
  {http://arxiv.org/abs/https://science.sciencemag.org/content/349/6243/47.full.pdf}
  {https://science.sciencemag.org/content/349/6243/47.full.pdf} \BibitemShut
  {NoStop}%
\bibitem [{\citenamefont {Lu}\ \emph {et~al.}(2014)\citenamefont {Lu},
  \citenamefont {Joannopoulos},\ and\ \citenamefont
  {Solja{\v{c}}i{\'c}}}]{luNPhoton2014}%
  \BibitemOpen
  \bibfield  {author} {\bibinfo {author} {\bibfnamefont {L.}~\bibnamefont
  {Lu}}, \bibinfo {author} {\bibfnamefont {J.~D.}\ \bibnamefont
  {Joannopoulos}}, \ and\ \bibinfo {author} {\bibfnamefont {M.}~\bibnamefont
  {Solja{\v{c}}i{\'c}}},\ }\href@noop {} {\bibfield  {journal} {\bibinfo
  {journal} {Nature Photonics}\ }\textbf {\bibinfo {volume} {8}},\ \bibinfo
  {pages} {821} (\bibinfo {year} {2014})}\BibitemShut {NoStop}%
\bibitem [{\citenamefont {Khanikaev}\ and\ \citenamefont
  {Shvets}(2017)}]{khanikaevNPhoton2017}%
  \BibitemOpen
  \bibfield  {author} {\bibinfo {author} {\bibfnamefont {A.~B.}\ \bibnamefont
  {Khanikaev}}\ and\ \bibinfo {author} {\bibfnamefont {G.}~\bibnamefont
  {Shvets}},\ }\href@noop {} {\bibfield  {journal} {\bibinfo  {journal} {Nature
  Photonics}\ }\textbf {\bibinfo {volume} {11}},\ \bibinfo {pages} {763}
  (\bibinfo {year} {2017})}\BibitemShut {NoStop}%
\bibitem [{\citenamefont {Rider}\ \emph {et~al.}(2019)\citenamefont {Rider},
  \citenamefont {Palmer}, \citenamefont {Pocock}, \citenamefont {Xiao},
  \citenamefont {Arroyo~Huidobro},\ and\ \citenamefont
  {Giannini}}]{riderJAP2019}%
  \BibitemOpen
  \bibfield  {author} {\bibinfo {author} {\bibfnamefont {M.~S.}\ \bibnamefont
  {Rider}}, \bibinfo {author} {\bibfnamefont {S.~J.}\ \bibnamefont {Palmer}},
  \bibinfo {author} {\bibfnamefont {S.~R.}\ \bibnamefont {Pocock}}, \bibinfo
  {author} {\bibfnamefont {X.}~\bibnamefont {Xiao}}, \bibinfo {author}
  {\bibfnamefont {P.}~\bibnamefont {Arroyo~Huidobro}}, \ and\ \bibinfo {author}
  {\bibfnamefont {V.}~\bibnamefont {Giannini}},\ }\href {\doibase
  10.1063/1.5086433} {\bibfield  {journal} {\bibinfo  {journal} {Journal of
  Applied Physics}\ }\textbf {\bibinfo {volume} {125}},\ \bibinfo {pages}
  {120901} (\bibinfo {year} {2019})},\ \Eprint
  {http://arxiv.org/abs/https://doi.org/10.1063/1.5086433}
  {https://doi.org/10.1063/1.5086433} \BibitemShut {NoStop}%
\bibitem [{\citenamefont {Xie}\ \emph {et~al.}(2018)\citenamefont {Xie},
  \citenamefont {Wang}, \citenamefont {Zhu}, \citenamefont {Lu}, \citenamefont
  {Wang},\ and\ \citenamefont {Chen}}]{xieOE2018}%
  \BibitemOpen
  \bibfield  {author} {\bibinfo {author} {\bibfnamefont {B.-Y.}\ \bibnamefont
  {Xie}}, \bibinfo {author} {\bibfnamefont {H.-F.}\ \bibnamefont {Wang}},
  \bibinfo {author} {\bibfnamefont {X.-Y.}\ \bibnamefont {Zhu}}, \bibinfo
  {author} {\bibfnamefont {M.-H.}\ \bibnamefont {Lu}}, \bibinfo {author}
  {\bibfnamefont {Z.~D.}\ \bibnamefont {Wang}}, \ and\ \bibinfo {author}
  {\bibfnamefont {Y.-F.}\ \bibnamefont {Chen}},\ }\href {\doibase
  10.1364/OE.26.024531} {\bibfield  {journal} {\bibinfo  {journal} {Opt.
  Express}\ }\textbf {\bibinfo {volume} {26}},\ \bibinfo {pages} {24531}
  (\bibinfo {year} {2018})}\BibitemShut {NoStop}%
\bibitem [{\citenamefont {Poli}\ \emph {et~al.}(2015)\citenamefont {Poli},
  \citenamefont {Bellec}, \citenamefont {Kuhl}, \citenamefont {Mortessagne},\
  and\ \citenamefont {Schomerus}}]{poliNComms2015}%
  \BibitemOpen
  \bibfield  {author} {\bibinfo {author} {\bibfnamefont {C.}~\bibnamefont
  {Poli}}, \bibinfo {author} {\bibfnamefont {M.}~\bibnamefont {Bellec}},
  \bibinfo {author} {\bibfnamefont {U.}~\bibnamefont {Kuhl}}, \bibinfo {author}
  {\bibfnamefont {F.}~\bibnamefont {Mortessagne}}, \ and\ \bibinfo {author}
  {\bibfnamefont {H.}~\bibnamefont {Schomerus}},\ }\href@noop {} {\bibfield
  {journal} {\bibinfo  {journal} {Nature communications}\ }\textbf {\bibinfo
  {volume} {6}},\ \bibinfo {pages} {6710} (\bibinfo {year} {2015})}\BibitemShut
  {NoStop}%
\bibitem [{\citenamefont {El-Ganainy}\ and\ \citenamefont
  {Levy}(2015)}]{el-GanainyOL2015}%
  \BibitemOpen
  \bibfield  {author} {\bibinfo {author} {\bibfnamefont {R.}~\bibnamefont
  {El-Ganainy}}\ and\ \bibinfo {author} {\bibfnamefont {M.}~\bibnamefont
  {Levy}},\ }\href {\doibase 10.1364/OL.40.005275} {\bibfield  {journal}
  {\bibinfo  {journal} {Opt. Lett.}\ }\textbf {\bibinfo {volume} {40}},\
  \bibinfo {pages} {5275} (\bibinfo {year} {2015})}\BibitemShut {NoStop}%
\bibitem [{\citenamefont {St-Jean}\ \emph {et~al.}(2017)\citenamefont
  {St-Jean}, \citenamefont {Goblot}, \citenamefont {Galopin}, \citenamefont
  {Lema{\^\i}tre}, \citenamefont {Ozawa}, \citenamefont {Le~Gratiet},
  \citenamefont {Sagnes}, \citenamefont {Bloch},\ and\ \citenamefont
  {Amo}}]{stjeanNaturephoton2017}%
  \BibitemOpen
  \bibfield  {author} {\bibinfo {author} {\bibfnamefont {P.}~\bibnamefont
  {St-Jean}}, \bibinfo {author} {\bibfnamefont {V.}~\bibnamefont {Goblot}},
  \bibinfo {author} {\bibfnamefont {E.}~\bibnamefont {Galopin}}, \bibinfo
  {author} {\bibfnamefont {A.}~\bibnamefont {Lema{\^\i}tre}}, \bibinfo {author}
  {\bibfnamefont {T.}~\bibnamefont {Ozawa}}, \bibinfo {author} {\bibfnamefont
  {L.}~\bibnamefont {Le~Gratiet}}, \bibinfo {author} {\bibfnamefont
  {I.}~\bibnamefont {Sagnes}}, \bibinfo {author} {\bibfnamefont
  {J.}~\bibnamefont {Bloch}}, \ and\ \bibinfo {author} {\bibfnamefont
  {A.}~\bibnamefont {Amo}},\ }\href {\doibase 10.1038/s41566-017-0006-2}
  {\bibfield  {journal} {\bibinfo  {journal} {Nature Photonics}\ }\textbf
  {\bibinfo {volume} {11}},\ \bibinfo {pages} {651} (\bibinfo {year}
  {2017})}\BibitemShut {NoStop}%
\bibitem [{\citenamefont {Parto}\ \emph {et~al.}(2018)\citenamefont {Parto},
  \citenamefont {Wittek}, \citenamefont {Hodaei}, \citenamefont {Harari},
  \citenamefont {Bandres}, \citenamefont {Ren}, \citenamefont {Rechtsman},
  \citenamefont {Segev}, \citenamefont {Christodoulides},\ and\ \citenamefont
  {Khajavikhan}}]{partoPRL2018}%
  \BibitemOpen
  \bibfield  {author} {\bibinfo {author} {\bibfnamefont {M.}~\bibnamefont
  {Parto}}, \bibinfo {author} {\bibfnamefont {S.}~\bibnamefont {Wittek}},
  \bibinfo {author} {\bibfnamefont {H.}~\bibnamefont {Hodaei}}, \bibinfo
  {author} {\bibfnamefont {G.}~\bibnamefont {Harari}}, \bibinfo {author}
  {\bibfnamefont {M.~A.}\ \bibnamefont {Bandres}}, \bibinfo {author}
  {\bibfnamefont {J.}~\bibnamefont {Ren}}, \bibinfo {author} {\bibfnamefont
  {M.~C.}\ \bibnamefont {Rechtsman}}, \bibinfo {author} {\bibfnamefont
  {M.}~\bibnamefont {Segev}}, \bibinfo {author} {\bibfnamefont {D.~N.}\
  \bibnamefont {Christodoulides}}, \ and\ \bibinfo {author} {\bibfnamefont
  {M.}~\bibnamefont {Khajavikhan}},\ }\href {\doibase
  10.1103/PhysRevLett.120.113901} {\bibfield  {journal} {\bibinfo  {journal}
  {Phys. Rev. Lett.}\ }\textbf {\bibinfo {volume} {120}},\ \bibinfo {pages}
  {113901} (\bibinfo {year} {2018})}\BibitemShut {NoStop}%
\bibitem [{\citenamefont {Zhao}\ \emph {et~al.}(2018)\citenamefont {Zhao},
  \citenamefont {Miao}, \citenamefont {Teimourpour}, \citenamefont {Malzard},
  \citenamefont {El-Ganainy}, \citenamefont {Schomerus},\ and\ \citenamefont
  {Feng}}]{zhaoNaturecomms2018}%
  \BibitemOpen
  \bibfield  {author} {\bibinfo {author} {\bibfnamefont {H.}~\bibnamefont
  {Zhao}}, \bibinfo {author} {\bibfnamefont {P.}~\bibnamefont {Miao}}, \bibinfo
  {author} {\bibfnamefont {M.~H.}\ \bibnamefont {Teimourpour}}, \bibinfo
  {author} {\bibfnamefont {S.}~\bibnamefont {Malzard}}, \bibinfo {author}
  {\bibfnamefont {R.}~\bibnamefont {El-Ganainy}}, \bibinfo {author}
  {\bibfnamefont {H.}~\bibnamefont {Schomerus}}, \ and\ \bibinfo {author}
  {\bibfnamefont {L.}~\bibnamefont {Feng}},\ }\href@noop {} {\bibfield
  {journal} {\bibinfo  {journal} {Nature communications}\ }\textbf {\bibinfo
  {volume} {9}},\ \bibinfo {pages} {981} (\bibinfo {year} {2018})}\BibitemShut
  {NoStop}%
\bibitem [{\citenamefont {Ling}\ \emph {et~al.}(2015)\citenamefont {Ling},
  \citenamefont {Xiao}, \citenamefont {Chan}, \citenamefont {Yu},\ and\
  \citenamefont {Fung}}]{lingOE2015}%
  \BibitemOpen
  \bibfield  {author} {\bibinfo {author} {\bibfnamefont {C.~W.}\ \bibnamefont
  {Ling}}, \bibinfo {author} {\bibfnamefont {M.}~\bibnamefont {Xiao}}, \bibinfo
  {author} {\bibfnamefont {C.~T.}\ \bibnamefont {Chan}}, \bibinfo {author}
  {\bibfnamefont {S.~F.}\ \bibnamefont {Yu}}, \ and\ \bibinfo {author}
  {\bibfnamefont {K.~H.}\ \bibnamefont {Fung}},\ }\href {\doibase
  10.1364/OE.23.002021} {\bibfield  {journal} {\bibinfo  {journal} {Opt.
  Express}\ }\textbf {\bibinfo {volume} {23}},\ \bibinfo {pages} {2021}
  (\bibinfo {year} {2015})}\BibitemShut {NoStop}%
\bibitem [{\citenamefont {Downing}\ and\ \citenamefont
  {Weick}(2017)}]{downingPRB2017}%
  \BibitemOpen
  \bibfield  {author} {\bibinfo {author} {\bibfnamefont {C.~A.}\ \bibnamefont
  {Downing}}\ and\ \bibinfo {author} {\bibfnamefont {G.}~\bibnamefont
  {Weick}},\ }\href {\doibase 10.1103/PhysRevB.95.125426} {\bibfield  {journal}
  {\bibinfo  {journal} {Phys. Rev. B}\ }\textbf {\bibinfo {volume} {95}},\
  \bibinfo {pages} {125426} (\bibinfo {year} {2017})}\BibitemShut {NoStop}%
\bibitem [{\citenamefont {Pocock}\ \emph {et~al.}(2018)\citenamefont {Pocock},
  \citenamefont {Xiao}, \citenamefont {Huidobro},\ and\ \citenamefont
  {Giannini}}]{pocockArxiv2017}%
  \BibitemOpen
  \bibfield  {author} {\bibinfo {author} {\bibfnamefont {S.}~\bibnamefont
  {Pocock}}, \bibinfo {author} {\bibfnamefont {X.}~\bibnamefont {Xiao}},
  \bibinfo {author} {\bibfnamefont {P.~A.}\ \bibnamefont {Huidobro}}, \ and\
  \bibinfo {author} {\bibfnamefont {V.}~\bibnamefont {Giannini}},\ }\href
  {\doibase 10.1021/acsphotonics.8b00117} {\bibfield  {journal} {\bibinfo
  {journal} {ACS Photonics}\ }\textbf {\bibinfo {volume} {5}},\ \bibinfo
  {pages} {2271} (\bibinfo {year} {2018})},\ \Eprint
  {http://arxiv.org/abs/https://doi.org/10.1021/acsphotonics.8b00117}
  {https://doi.org/10.1021/acsphotonics.8b00117} \BibitemShut {NoStop}%
\bibitem [{\citenamefont {Downing}\ and\ \citenamefont
  {Weick}(2018)}]{downing2018topological}%
  \BibitemOpen
  \bibfield  {author} {\bibinfo {author} {\bibfnamefont {C.~A.}\ \bibnamefont
  {Downing}}\ and\ \bibinfo {author} {\bibfnamefont {G.}~\bibnamefont
  {Weick}},\ }\href {\doibase 10.1140/epjb/e2018-90199-0} {\bibfield  {journal}
  {\bibinfo  {journal} {The European Physical Journal B}\ }\textbf {\bibinfo
  {volume} {91}},\ \bibinfo {pages} {253} (\bibinfo {year} {2018})}\BibitemShut
  {NoStop}%
\bibitem [{\citenamefont {Pocock}\ \emph {et~al.}(2019)\citenamefont {Pocock},
  \citenamefont {Huidobro},\ and\ \citenamefont
  {Vincenzo}}]{pocockNanophoton2019}%
  \BibitemOpen
  \bibfield  {author} {\bibinfo {author} {\bibfnamefont {S.~R.}\ \bibnamefont
  {Pocock}}, \bibinfo {author} {\bibfnamefont {P.~A.}\ \bibnamefont
  {Huidobro}}, \ and\ \bibinfo {author} {\bibfnamefont {G.}~\bibnamefont
  {Vincenzo}},\ }\href {\doibase 10.1515/nanoph-2019-0033} {\bibfield
  {journal} {\bibinfo  {journal} {Nanophotonics}\ }\textbf {\bibinfo {volume}
  {8}},\ \bibinfo {pages} {1337} (\bibinfo {year} {2019})}\BibitemShut
  {NoStop}%
\bibitem [{\citenamefont {Xu}\ \emph {et~al.}(2019)\citenamefont {Xu},
  \citenamefont {Zhang}, \citenamefont {Zhao}, \citenamefont {Guo},
  \citenamefont {Huang},\ and\ \citenamefont {Ke}}]{xuAppsci2019}%
  \BibitemOpen
  \bibfield  {author} {\bibinfo {author} {\bibfnamefont {C.}~\bibnamefont
  {Xu}}, \bibinfo {author} {\bibfnamefont {P.}~\bibnamefont {Zhang}}, \bibinfo
  {author} {\bibfnamefont {D.}~\bibnamefont {Zhao}}, \bibinfo {author}
  {\bibfnamefont {H.}~\bibnamefont {Guo}}, \bibinfo {author} {\bibfnamefont
  {M.}~\bibnamefont {Huang}}, \ and\ \bibinfo {author} {\bibfnamefont
  {S.}~\bibnamefont {Ke}},\ }\href {\doibase 10.3390/app9194152} {\bibfield
  {journal} {\bibinfo  {journal} {Applied Sciences}\ }\textbf {\bibinfo
  {volume} {9}} (\bibinfo {year} {2019}),\ 10.3390/app9194152}\BibitemShut
  {NoStop}%
\bibitem [{\citenamefont {{You}}\ \emph {et~al.}(2019)\citenamefont {{You}},
  \citenamefont {{Lan}},\ and\ \citenamefont {{Panoiu}}}]{you2019four-wave}%
  \BibitemOpen
  \bibfield  {author} {\bibinfo {author} {\bibfnamefont {J.~W.}\ \bibnamefont
  {{You}}}, \bibinfo {author} {\bibfnamefont {Z.}~\bibnamefont {{Lan}}}, \ and\
  \bibinfo {author} {\bibfnamefont {N.~C.}\ \bibnamefont {{Panoiu}}},\
  }\href@noop {} {\bibfield  {journal} {\bibinfo  {journal} {arXiv e-prints}\
  ,\ \bibinfo {eid} {arXiv:1908.05477}} (\bibinfo {year} {2019})},\ \Eprint
  {http://arxiv.org/abs/1908.05477} {arXiv:1908.05477 [physics.optics]}
  \BibitemShut {NoStop}%
\bibitem [{\citenamefont {Su}\ \emph {et~al.}(1979)\citenamefont {Su},
  \citenamefont {Schrieffer},\ and\ \citenamefont {Heeger}}]{suPRL1979}%
  \BibitemOpen
  \bibfield  {author} {\bibinfo {author} {\bibfnamefont {W.~P.}\ \bibnamefont
  {Su}}, \bibinfo {author} {\bibfnamefont {J.~R.}\ \bibnamefont {Schrieffer}},
  \ and\ \bibinfo {author} {\bibfnamefont {A.~J.}\ \bibnamefont {Heeger}},\
  }\href {\doibase 10.1103/PhysRevLett.42.1698} {\bibfield  {journal} {\bibinfo
   {journal} {Phys. Rev. Lett.}\ }\textbf {\bibinfo {volume} {42}},\ \bibinfo
  {pages} {1698} (\bibinfo {year} {1979})}\BibitemShut {NoStop}%
\bibitem [{\citenamefont {Halevi}\ and\ \citenamefont
  {Ramos-Mendieta}(2000)}]{haleviPRL2000}%
  \BibitemOpen
  \bibfield  {author} {\bibinfo {author} {\bibfnamefont {P.}~\bibnamefont
  {Halevi}}\ and\ \bibinfo {author} {\bibfnamefont {F.}~\bibnamefont
  {Ramos-Mendieta}},\ }\href {\doibase 10.1103/PhysRevLett.85.1875} {\bibfield
  {journal} {\bibinfo  {journal} {Phys. Rev. Lett.}\ }\textbf {\bibinfo
  {volume} {85}},\ \bibinfo {pages} {1875} (\bibinfo {year}
  {2000})}\BibitemShut {NoStop}%
\bibitem [{\citenamefont {G\'omez~Rivas}\ \emph {et~al.}(2006)\citenamefont
  {G\'omez~Rivas}, \citenamefont {S\'anchez-Gil}, \citenamefont {Kuttge},
  \citenamefont {Haring~Bolivar},\ and\ \citenamefont
  {Kurz}}]{gomezrivasPRB2006}%
  \BibitemOpen
  \bibfield  {author} {\bibinfo {author} {\bibfnamefont {J.}~\bibnamefont
  {G\'omez~Rivas}}, \bibinfo {author} {\bibfnamefont {J.~A.}\ \bibnamefont
  {S\'anchez-Gil}}, \bibinfo {author} {\bibfnamefont {M.}~\bibnamefont
  {Kuttge}}, \bibinfo {author} {\bibfnamefont {P.}~\bibnamefont
  {Haring~Bolivar}}, \ and\ \bibinfo {author} {\bibfnamefont {H.}~\bibnamefont
  {Kurz}},\ }\href {\doibase 10.1103/PhysRevB.74.245324} {\bibfield  {journal}
  {\bibinfo  {journal} {Phys. Rev. B}\ }\textbf {\bibinfo {volume} {74}},\
  \bibinfo {pages} {245324} (\bibinfo {year} {2006})}\BibitemShut {NoStop}%
\bibitem [{\citenamefont {Wang}\ and\ \citenamefont
  {Zhao}(2018{\natexlab{a}})}]{wang2018topological}%
  \BibitemOpen
  \bibfield  {author} {\bibinfo {author} {\bibfnamefont {B.~X.}\ \bibnamefont
  {Wang}}\ and\ \bibinfo {author} {\bibfnamefont {C.~Y.}\ \bibnamefont
  {Zhao}},\ }\href {\doibase 10.1103/PhysRevA.98.023808} {\bibfield  {journal}
  {\bibinfo  {journal} {Phys. Rev. A}\ }\textbf {\bibinfo {volume} {98}},\
  \bibinfo {pages} {023808} (\bibinfo {year} {2018}{\natexlab{a}})}\BibitemShut
  {NoStop}%
\bibitem [{\citenamefont {Wang}\ and\ \citenamefont
  {Zhao}(2018{\natexlab{b}})}]{wangPRB2018b}%
  \BibitemOpen
  \bibfield  {author} {\bibinfo {author} {\bibfnamefont {B.~X.}\ \bibnamefont
  {Wang}}\ and\ \bibinfo {author} {\bibfnamefont {C.~Y.}\ \bibnamefont
  {Zhao}},\ }\href {\doibase 10.1103/PhysRevB.98.165435} {\bibfield  {journal}
  {\bibinfo  {journal} {Phys. Rev. B}\ }\textbf {\bibinfo {volume} {98}},\
  \bibinfo {pages} {165435} (\bibinfo {year} {2018}{\natexlab{b}})}\BibitemShut
  {NoStop}%
\bibitem [{\citenamefont {Tervo}\ \emph {et~al.}(2017)\citenamefont {Tervo},
  \citenamefont {Zhang},\ and\ \citenamefont {Cola}}]{tervoPRMater2018}%
  \BibitemOpen
  \bibfield  {author} {\bibinfo {author} {\bibfnamefont {E.}~\bibnamefont
  {Tervo}}, \bibinfo {author} {\bibfnamefont {Z.}~\bibnamefont {Zhang}}, \ and\
  \bibinfo {author} {\bibfnamefont {B.}~\bibnamefont {Cola}},\ }\href {\doibase
  10.1103/PhysRevMaterials.1.015201} {\bibfield  {journal} {\bibinfo  {journal}
  {Phys. Rev. Materials}\ }\textbf {\bibinfo {volume} {1}},\ \bibinfo {pages}
  {015201} (\bibinfo {year} {2017})}\BibitemShut {NoStop}%
\bibitem [{\citenamefont {Markel}\ and\ \citenamefont
  {Sarychev}(2007)}]{markelPRB2007}%
  \BibitemOpen
  \bibfield  {author} {\bibinfo {author} {\bibfnamefont {V.~A.}\ \bibnamefont
  {Markel}}\ and\ \bibinfo {author} {\bibfnamefont {A.~K.}\ \bibnamefont
  {Sarychev}},\ }\href {\doibase 10.1103/PhysRevB.75.085426} {\bibfield
  {journal} {\bibinfo  {journal} {Phys. Rev. B}\ }\textbf {\bibinfo {volume}
  {75}},\ \bibinfo {pages} {085426} (\bibinfo {year} {2007})}\BibitemShut
  {NoStop}%
\bibitem [{\citenamefont {Park}\ and\ \citenamefont
  {Stroud}(2004)}]{parkPRB2004}%
  \BibitemOpen
  \bibfield  {author} {\bibinfo {author} {\bibfnamefont {S.~Y.}\ \bibnamefont
  {Park}}\ and\ \bibinfo {author} {\bibfnamefont {D.}~\bibnamefont {Stroud}},\
  }\href {\doibase 10.1103/PhysRevB.69.125418} {\bibfield  {journal} {\bibinfo
  {journal} {Phys. Rev. B}\ }\textbf {\bibinfo {volume} {69}},\ \bibinfo
  {pages} {125418} (\bibinfo {year} {2004})}\BibitemShut {NoStop}%
\bibitem [{\citenamefont {Cunningham}\ and\ \citenamefont
  {Gruber}(1970)}]{cunninghamJAP1970}%
  \BibitemOpen
  \bibfield  {author} {\bibinfo {author} {\bibfnamefont {R.~W.}\ \bibnamefont
  {Cunningham}}\ and\ \bibinfo {author} {\bibfnamefont {J.~B.}\ \bibnamefont
  {Gruber}},\ }\href {\doibase 10.1063/1.1659107} {\bibfield  {journal}
  {\bibinfo  {journal} {Journal of Applied Physics}\ }\textbf {\bibinfo
  {volume} {41}},\ \bibinfo {pages} {1804} (\bibinfo {year} {1970})},\ \Eprint
  {http://arxiv.org/abs/https://doi.org/10.1063/1.1659107}
  {https://doi.org/10.1063/1.1659107} \BibitemShut {NoStop}%
\bibitem [{\citenamefont {Han}\ and\ \citenamefont
  {Lakhtakia}(2009)}]{hanJMO2009}%
  \BibitemOpen
  \bibfield  {author} {\bibinfo {author} {\bibfnamefont {J.}~\bibnamefont
  {Han}}\ and\ \bibinfo {author} {\bibfnamefont {A.}~\bibnamefont
  {Lakhtakia}},\ }\href {\doibase 10.1080/09500340802621785} {\bibfield
  {journal} {\bibinfo  {journal} {Journal of Modern Optics}\ }\textbf {\bibinfo
  {volume} {56}},\ \bibinfo {pages} {554} (\bibinfo {year} {2009})},\ \Eprint
  {http://arxiv.org/abs/https://doi.org/10.1080/09500340802621785}
  {https://doi.org/10.1080/09500340802621785} \BibitemShut {NoStop}%
\bibitem [{\citenamefont {Howells}\ and\ \citenamefont
  {Schlie}(1996)}]{howellsAPL1996}%
  \BibitemOpen
  \bibfield  {author} {\bibinfo {author} {\bibfnamefont {S.~C.}\ \bibnamefont
  {Howells}}\ and\ \bibinfo {author} {\bibfnamefont {L.~A.}\ \bibnamefont
  {Schlie}},\ }\href {\doibase 10.1063/1.117783} {\bibfield  {journal}
  {\bibinfo  {journal} {Applied Physics Letters}\ }\textbf {\bibinfo {volume}
  {69}},\ \bibinfo {pages} {550} (\bibinfo {year} {1996})},\ \Eprint
  {http://arxiv.org/abs/https://doi.org/10.1063/1.117783}
  {https://doi.org/10.1063/1.117783} \BibitemShut {NoStop}%
\bibitem [{\citenamefont {Kittel}\ \emph {et~al.}(1976)\citenamefont {Kittel}
  \emph {et~al.}}]{kittel1976}%
  \BibitemOpen
  \bibfield  {author} {\bibinfo {author} {\bibfnamefont {C.}~\bibnamefont
  {Kittel}} \emph {et~al.},\ }\href@noop {} {\emph {\bibinfo {title}
  {Introduction to solid state physics}}},\ Vol.~\bibinfo {volume} {8}\
  (\bibinfo  {publisher} {Wiley New York},\ \bibinfo {year} {1976})\BibitemShut
  {NoStop}%
\bibitem [{\citenamefont {Weber}\ and\ \citenamefont
  {Ford}(2004)}]{weberPRB2004}%
  \BibitemOpen
  \bibfield  {author} {\bibinfo {author} {\bibfnamefont {W.~H.}\ \bibnamefont
  {Weber}}\ and\ \bibinfo {author} {\bibfnamefont {G.~W.}\ \bibnamefont
  {Ford}},\ }\href {\doibase 10.1103/PhysRevB.70.125429} {\bibfield  {journal}
  {\bibinfo  {journal} {Phys. Rev. B}\ }\textbf {\bibinfo {volume} {70}},\
  \bibinfo {pages} {125429} (\bibinfo {year} {2004})}\BibitemShut {NoStop}%
\bibitem [{\citenamefont {Lieu}(2018)}]{lieuPRB2018}%
  \BibitemOpen
  \bibfield  {author} {\bibinfo {author} {\bibfnamefont {S.}~\bibnamefont
  {Lieu}},\ }\href {\doibase 10.1103/PhysRevB.97.045106} {\bibfield  {journal}
  {\bibinfo  {journal} {Phys. Rev. B}\ }\textbf {\bibinfo {volume} {97}},\
  \bibinfo {pages} {045106} (\bibinfo {year} {2018})}\BibitemShut {NoStop}%
\bibitem [{\citenamefont {Wang}\ \emph {et~al.}(2018)\citenamefont {Wang},
  \citenamefont {R\"{o}ntgen}, \citenamefont {Morfonios}, \citenamefont
  {Pinheiro}, \citenamefont {Schmelcher},\ and\ \citenamefont
  {Negro}}]{wangOL2018}%
  \BibitemOpen
  \bibfield  {author} {\bibinfo {author} {\bibfnamefont {R.}~\bibnamefont
  {Wang}}, \bibinfo {author} {\bibfnamefont {M.}~\bibnamefont {R\"{o}ntgen}},
  \bibinfo {author} {\bibfnamefont {C.~V.}\ \bibnamefont {Morfonios}}, \bibinfo
  {author} {\bibfnamefont {F.~A.}\ \bibnamefont {Pinheiro}}, \bibinfo {author}
  {\bibfnamefont {P.}~\bibnamefont {Schmelcher}}, \ and\ \bibinfo {author}
  {\bibfnamefont {L.~D.}\ \bibnamefont {Negro}},\ }\href {\doibase
  10.1364/OL.43.001986} {\bibfield  {journal} {\bibinfo  {journal} {Opt.
  Lett.}\ }\textbf {\bibinfo {volume} {43}},\ \bibinfo {pages} {1986} (\bibinfo
  {year} {2018})}\BibitemShut {NoStop}%
\bibitem [{\citenamefont {{Ma}}\ \emph {et~al.}(2017)\citenamefont {{Ma}},
  \citenamefont {{Nguyen-Huu}}, \citenamefont {{Zhou}}, \citenamefont
  {{Maeda}}, \citenamefont {{Wu}}, \citenamefont {{Eldlio}}, \citenamefont
  {{Pištora}},\ and\ \citenamefont {{Cada}}}]{maIEEEJSTQE2017}%
  \BibitemOpen
  \bibfield  {author} {\bibinfo {author} {\bibfnamefont {Y.}~\bibnamefont
  {{Ma}}}, \bibinfo {author} {\bibfnamefont {N.}~\bibnamefont {{Nguyen-Huu}}},
  \bibinfo {author} {\bibfnamefont {J.}~\bibnamefont {{Zhou}}}, \bibinfo
  {author} {\bibfnamefont {H.}~\bibnamefont {{Maeda}}}, \bibinfo {author}
  {\bibfnamefont {Q.}~\bibnamefont {{Wu}}}, \bibinfo {author} {\bibfnamefont
  {M.}~\bibnamefont {{Eldlio}}}, \bibinfo {author} {\bibfnamefont
  {J.}~\bibnamefont {{Pištora}}}, \ and\ \bibinfo {author} {\bibfnamefont
  {M.}~\bibnamefont {{Cada}}},\ }\href {\doibase 10.1109/JSTQE.2017.2660882}
  {\bibfield  {journal} {\bibinfo  {journal} {IEEE Journal of Selected Topics
  in Quantum Electronics}\ }\textbf {\bibinfo {volume} {23}},\ \bibinfo {pages}
  {1} (\bibinfo {year} {2017})}\BibitemShut {NoStop}%
\bibitem [{\citenamefont {Proctor}\ \emph {et~al.}(0)\citenamefont {Proctor},
  \citenamefont {Craster}, \citenamefont {Maier}, \citenamefont {Giannini},\
  and\ \citenamefont {Huidobro}}]{proctor2019exciting}%
  \BibitemOpen
  \bibfield  {author} {\bibinfo {author} {\bibfnamefont {M.}~\bibnamefont
  {Proctor}}, \bibinfo {author} {\bibfnamefont {R.~V.}\ \bibnamefont
  {Craster}}, \bibinfo {author} {\bibfnamefont {S.~A.}\ \bibnamefont {Maier}},
  \bibinfo {author} {\bibfnamefont {V.}~\bibnamefont {Giannini}}, \ and\
  \bibinfo {author} {\bibfnamefont {P.~A.}\ \bibnamefont {Huidobro}},\ }\href
  {\doibase 10.1021/acsphotonics.9b01192} {\bibfield  {journal} {\bibinfo
  {journal} {ACS Photonics}\ }\textbf {\bibinfo {volume} {0}},\ \bibinfo
  {pages} {null} (\bibinfo {year} {0})},\ \Eprint
  {http://arxiv.org/abs/https://doi.org/10.1021/acsphotonics.9b01192}
  {https://doi.org/10.1021/acsphotonics.9b01192} \BibitemShut {NoStop}%
\bibitem [{\citenamefont {Slobozhanyuk}\ \emph {et~al.}(2019)\citenamefont
  {Slobozhanyuk}, \citenamefont {Shchelokova}, \citenamefont {Ni},
  \citenamefont {Hossein~Mousavi}, \citenamefont {Smirnova}, \citenamefont
  {Belov}, \citenamefont {Alù}, \citenamefont {Kivshar},\ and\ \citenamefont
  {Khanikaev}}]{slobozhanyukAPL2019}%
  \BibitemOpen
  \bibfield  {author} {\bibinfo {author} {\bibfnamefont {A.}~\bibnamefont
  {Slobozhanyuk}}, \bibinfo {author} {\bibfnamefont {A.~V.}\ \bibnamefont
  {Shchelokova}}, \bibinfo {author} {\bibfnamefont {X.}~\bibnamefont {Ni}},
  \bibinfo {author} {\bibfnamefont {S.}~\bibnamefont {Hossein~Mousavi}},
  \bibinfo {author} {\bibfnamefont {D.~A.}\ \bibnamefont {Smirnova}}, \bibinfo
  {author} {\bibfnamefont {P.~A.}\ \bibnamefont {Belov}}, \bibinfo {author}
  {\bibfnamefont {A.}~\bibnamefont {Alù}}, \bibinfo {author} {\bibfnamefont
  {Y.~S.}\ \bibnamefont {Kivshar}}, \ and\ \bibinfo {author} {\bibfnamefont
  {A.~B.}\ \bibnamefont {Khanikaev}},\ }\href {\doibase 10.1063/1.5055601}
  {\bibfield  {journal} {\bibinfo  {journal} {Applied Physics Letters}\
  }\textbf {\bibinfo {volume} {114}},\ \bibinfo {pages} {031103} (\bibinfo
  {year} {2019})},\ \Eprint
  {http://arxiv.org/abs/https://doi.org/10.1063/1.5055601}
  {https://doi.org/10.1063/1.5055601} \BibitemShut {NoStop}%
\bibitem [{\citenamefont {Peng}\ \emph {et~al.}(2019)\citenamefont {Peng},
  \citenamefont {Schilder}, \citenamefont {Ni}, \citenamefont {van~de Groep},
  \citenamefont {Brongersma}, \citenamefont {Al\`u}, \citenamefont {Khanikaev},
  \citenamefont {Atwater},\ and\ \citenamefont {Polman}}]{pengPRL2019}%
  \BibitemOpen
  \bibfield  {author} {\bibinfo {author} {\bibfnamefont {S.}~\bibnamefont
  {Peng}}, \bibinfo {author} {\bibfnamefont {N.~J.}\ \bibnamefont {Schilder}},
  \bibinfo {author} {\bibfnamefont {X.}~\bibnamefont {Ni}}, \bibinfo {author}
  {\bibfnamefont {J.}~\bibnamefont {van~de Groep}}, \bibinfo {author}
  {\bibfnamefont {M.~L.}\ \bibnamefont {Brongersma}}, \bibinfo {author}
  {\bibfnamefont {A.}~\bibnamefont {Al\`u}}, \bibinfo {author} {\bibfnamefont
  {A.~B.}\ \bibnamefont {Khanikaev}}, \bibinfo {author} {\bibfnamefont {H.~A.}\
  \bibnamefont {Atwater}}, \ and\ \bibinfo {author} {\bibfnamefont
  {A.}~\bibnamefont {Polman}},\ }\href {\doibase
  10.1103/PhysRevLett.122.117401} {\bibfield  {journal} {\bibinfo  {journal}
  {Phys. Rev. Lett.}\ }\textbf {\bibinfo {volume} {122}},\ \bibinfo {pages}
  {117401} (\bibinfo {year} {2019})}\BibitemShut {NoStop}%
\bibitem [{\citenamefont {Wu}\ \emph {et~al.}(2019)\citenamefont {Wu},
  \citenamefont {Li}, \citenamefont {Chen}, \citenamefont {Sheng},
  \citenamefont {Jing}, \citenamefont {Fan},\ and\ \citenamefont
  {Peng}}]{wuACSANM2019}%
  \BibitemOpen
  \bibfield  {author} {\bibinfo {author} {\bibfnamefont {H.-W.}\ \bibnamefont
  {Wu}}, \bibinfo {author} {\bibfnamefont {Y.}~\bibnamefont {Li}}, \bibinfo
  {author} {\bibfnamefont {H.-J.}\ \bibnamefont {Chen}}, \bibinfo {author}
  {\bibfnamefont {Z.-Q.}\ \bibnamefont {Sheng}}, \bibinfo {author}
  {\bibfnamefont {H.}~\bibnamefont {Jing}}, \bibinfo {author} {\bibfnamefont
  {R.-H.}\ \bibnamefont {Fan}}, \ and\ \bibinfo {author} {\bibfnamefont
  {R.-W.}\ \bibnamefont {Peng}},\ }\href {\doibase 10.1021/acsanm.8b02318}
  {\bibfield  {journal} {\bibinfo  {journal} {ACS Applied Nano Materials}\
  }\textbf {\bibinfo {volume} {2}},\ \bibinfo {pages} {1045} (\bibinfo {year}
  {2019})},\ \Eprint
  {http://arxiv.org/abs/https://doi.org/10.1021/acsanm.8b02318}
  {https://doi.org/10.1021/acsanm.8b02318} \BibitemShut {NoStop}%
\bibitem [{\citenamefont {Nguyen}\ \emph {et~al.}(2018)\citenamefont {Nguyen},
  \citenamefont {Evans}, \citenamefont {Sipahigil}, \citenamefont {Bhaskar},
  \citenamefont {Sukachev}, \citenamefont {Agafonov}, \citenamefont {Davydov},
  \citenamefont {Kulikova}, \citenamefont {Jelezko},\ and\ \citenamefont
  {Lukin}}]{nguyenAPL2018}%
  \BibitemOpen
  \bibfield  {author} {\bibinfo {author} {\bibfnamefont {C.~T.}\ \bibnamefont
  {Nguyen}}, \bibinfo {author} {\bibfnamefont {R.~E.}\ \bibnamefont {Evans}},
  \bibinfo {author} {\bibfnamefont {A.}~\bibnamefont {Sipahigil}}, \bibinfo
  {author} {\bibfnamefont {M.~K.}\ \bibnamefont {Bhaskar}}, \bibinfo {author}
  {\bibfnamefont {D.~D.}\ \bibnamefont {Sukachev}}, \bibinfo {author}
  {\bibfnamefont {V.~N.}\ \bibnamefont {Agafonov}}, \bibinfo {author}
  {\bibfnamefont {V.~A.}\ \bibnamefont {Davydov}}, \bibinfo {author}
  {\bibfnamefont {L.~F.}\ \bibnamefont {Kulikova}}, \bibinfo {author}
  {\bibfnamefont {F.}~\bibnamefont {Jelezko}}, \ and\ \bibinfo {author}
  {\bibfnamefont {M.~D.}\ \bibnamefont {Lukin}},\ }\href {\doibase
  10.1063/1.5029904} {\bibfield  {journal} {\bibinfo  {journal} {Applied
  Physics Letters}\ }\textbf {\bibinfo {volume} {112}},\ \bibinfo {pages}
  {203102} (\bibinfo {year} {2018})},\ \Eprint
  {http://arxiv.org/abs/https://doi.org/10.1063/1.5029904}
  {https://doi.org/10.1063/1.5029904} \BibitemShut {NoStop}%
\end{thebibliography}%


%

\end{document}